\documentclass[10pt]{article}

\usepackage{amsmath}

\usepackage{array}

\usepackage{appendix}

\usepackage{tocloft}                   

\usepackage{graphicx}

\usepackage{amsfonts}

\usepackage{amssymb}

\usepackage{mathrsfs}

\usepackage{yfonts}

\usepackage{euscript}

\usepackage{centernot}                 

\usepackage{ifsym}                     

\usepackage{upgreek}

\usepackage{mathtools}

\usepackage{bm}

\usepackage{color}

\usepackage{slantsc}
\usepackage{calligra}

\usepackage{bbold}          

\usepackage[T1]{fontenc}

\usepackage{epsf}

\usepackage{latexsym}

\usepackage{tipa}

\usepackage{makeidx}

\makeindex

\def\be{\begin{equation}}

\def\ee{\end{equation}}

\def\beq{\begin{equation}}

\def\eeq{\end{equation}}

\def\bea{\begin{eqnarray}}

\def\eea{\end{eqnarray}}

\def\!{\hspace{-1.6667em}}

\def\m{\mbox{ }}

\def\mma {\m , \m \m }

\def\!{\hspace{-1.6667em}}

\def\n{\noindent}

\def\u{\underline}

\def\es{\m = \m}

\def\:={\m := \m}

\def\=:{\m =: \m}

\def\bip{\mbox{\boldmath$p$}}

\def\biq{\mbox{\boldmath$q$}}

\def\sbiq{\mbox{\scriptsize\boldmath$q$}}

\def\sbiQ{\mbox{\scriptsize\boldmath$Q$}}

\def\sbiO{\mbox{\scriptsize\boldmath$O$}}

\def\sbiD{\mbox{\scriptsize\boldmath$D$}}

\def\sbiK{\mbox{\scriptsize\boldmath$K$}}

\def\sbiG{\mbox{\scriptsize\boldmath$G$}}

\def\sbiS{\mbox{\scriptsize\boldmath$S$}}

\def\sbiU{\mbox{\scriptsize\boldmath$U$}}

\def\tbiQ{\mbox{\tiny\boldmath$Q$}}

\def\tbiO{\mbox{\tiny\boldmath$O$}}

\def\tbiD{\mbox{\tiny\boldmath$D$}}

\def\tbiG{\mbox{\tiny\boldmath$G$}}

\def\tbiU{\mbox{\tiny\boldmath$U$}}

\def\bnabla{\mbox{\boldmath$\nabla$}}               
 
\def\birho{\mbox{\boldmath$\rho$}}

\def\brho{\birho}                                   

\def\bTheta{\mbox{\boldmath$\Theta$}} 

\def\mC{\mbox{C}}                        

\def\mb{\mbox{b}}

\def\mo{\mbox{o}}

\def\mt{\mbox{t}}

\def\uq{\underline{\mbox{$q$}}}

\def\ux{\underline{\mbox{x}}}

\def\bG{\mbox{\bf G}}                     

\def\btheta{\mbox{\boldmath$\theta$}}             

\def\bsigma{\mbox{\boldmath$\sigma$}}                   %

\def\sbcF{\mbox{\boldmath \scriptsize ${\cal F}$}}

\def\sbcS{\mbox{\boldmath \scriptsize ${\cal S}$}}

\def\tbcF{\mbox{\boldmath \tiny ${\cal F}$}}

\def\tbcL{\mbox{\boldmath \tiny ${\cal L}$}}

\def\tbcS{\mbox{\boldmath \tiny ${\cal S}$}}

\def\barr{\bar{r}}

\def\cr{\mbox{\scriptsize\bm{ $\mbox{ } \times \mbox{ }$}}}

\def\sumi2{\sum\mbox{}_{\mbox{}_{\mbox{\scriptsize $i$=1}}}^2}

\def\sumi3{\sum\mbox{}_{\mbox{}_{\mbox{\scriptsize $i$=1}}}^3}

\def\sumABcycles3{\sum\mbox{}_{\mbox{}_{\mbox{\scriptsize cycles $A,B$=1}}}^{3}}

\def\sumCDcycles3{\sum\mbox{}_{\mbox{}_{\mbox{\scriptsize cycles $C,D$=1}}}^{3}}

\def\sumj3{\sum\mbox{}_{\mbox{}_{\mbox{\scriptsize $j$=1}}}^3}

\def\sumk3{\sum\mbox{}_{\mbox{}_{\mbox{\scriptsize $k$=1}}}^3}

\def\prodiA1{\prod\mbox{}_{\mbox{}_{\mbox{\scriptsize $i$=1}}}^{A - 1}}

\def\bigtimes{\mbox{\Large $\times$}}

\def\d{\textrm{d}}                                                  

\def\pa{\partial}                                                   

\def\siF{\mbox{\scriptsize $F$}}

\def\siQ{\mbox{\scriptsize $Q$}}

\def\FrS{\mbox{\Large $\mathfrak{s}$}}                         
\def\FrD{\mbox{$\mathfrak{D}$}}	                               

\def\lFrg{\mbox{\Large$\mathfrak{g}$}}                         
\def\nFrg{\mbox{\large$\mathfrak{g}$}}                         

\def\Hilb{\mbox{{\boldmath$\mathfrak{H}$}ilb}}                 
\def\bFrW{\mbox{\boldmath$\mathfrak{W}$}} 					   

\def\scC{\mbox{\scriptsize ${\cal C}$}}                    
\def\scF{\mbox{\scriptsize ${\cal F}$}}

\def\scG{\mbox{\scriptsize ${\cal G}$}}                    

\def\scL{\mbox{\scriptsize ${\cal L}$}}                    

\def\scS{\mbox{\scriptsize ${\cal S}$}}                    

\def\siO{\mbox{\scriptsize$O$}}                             %

\def\FrQ{\mbox{\Large $\mathfrak{q}$}}                               
\def\Phase{\mbox{{\boldmath$\mathfrak{P}$}hase}}                     

\def\bFrR{\mbox{\boldmath$\mathfrak{R}$}}                            
\def\Rig-Phase{\bFrR\mbox{ig-}\Phase}                                
                                                                													   
\def\FrP{\mbox{\Large $\mathfrak{p}$}}                                 
\def\FrR{\mbox{\boldmath$\mathfrak{R}$}}                             

\def\bFrM{\mbox{\boldmath${\mathfrak{M}}$}}                             

\def\Positive-Modespace{\mbox{{\boldmath$\mathfrak{M}$}odespace$^+$}}

\def\POSITIVE-MODESPACE{\mbox{{\boldmath$\mathfrak{M}$}ODESPACE$^+$}}

\def\Kin-Hilb{\mbox{{\boldmath$\mathfrak{K}$}in-\Hilb}}                     

\def\Mid-Hilb{\mbox{{\boldmath$\mathfrak{M}$}id-\Hilb}}                     

\def\Dyn-Hilb{\mbox{{\boldmath$\mathfrak{D}$}yn-\Hilb}}                     

\def\5Star{\mbox{\Large$\star$}}              

\def\Frr{\mbox{$\mathfrak{r}$}}

\begin{document}

\begin{titlepage}

\begin{center}

\vspace{0.1in}

\Large{\bf Specific PDEs for Preserved Quantities in Geometry.} 

\vspace{0.1in}

\Large{\bf I. Similarities and Subgroups.} \normalsize

\vspace{0.1in}

{\large \bf Edward Anderson$^*$}

\end{center}

\begin{abstract}

We provide specific PDEs for preserved quantities $\sbiQ$ in Geometry, as well as a bridge between this and specific PDEs for observables $\sbiO$ in Physics.
We furthermore prove versions of four other theorems either side of this bridge: the below enumerated sentences.   
For the generic geometry - in the sense of it possessing no generalized Killing vectors, i.e.\ continuous geometrical automorphisms - 
the $\sbiQ$ form a smooth space of free functions over said geometry.
If a geometry possesses the corresponding type of Killing vectors, the $\sbiQ$ must Lie-brackets commute with `sums-over-points of the automorphism generators', $\sbcS$.
The observables counterpart of this is that in the presence of first-class constraints $\sbcF$, the $\sbiO$ must Poisson-brackets commute with these.  
Then 
1) defining $\sbiQ$, $\sbiO$ requires closed subalgebras of $\sbcS$, $\sbcF$.   
2) The $\sbiQ$, and the $\sbiO$, themselves form closed algebras. 
3) The subalgebras of $\sbiQ$, $\sbiO$  form bounded lattices dual to those of $\sbcS$, $\sbcF$ respectively. 
Both $\sbcS$, $\sbiQ$ and $\sbcF$, $\sbiO$ commutations can moreover be reformulated as first-order linear PDEs, treated free-characteristically.  
The secondmost generic case has just one $\sbcS$ or $\sbcF$, and so just one PDE, which standardly reduces to an ODE system.
The more highly nongeneric case of multiple $\sbcS$ or $\sbcF$, however, returns an over-determined PDE system.
4) We prove that nonetheless these are always integrable. 
This is significant by being mostly-opposite to how the more familiar generalized Killing equations themselves behave.  
We finally solve for the preserved quantities of similarity geometry and its subgroups; companion papers extend this program to affine, projective and conformal geometries. 

\end{abstract}

\n Mathematics keywords: Geometrical automorphism groups and the corresponding preserved quantities.  
Geometrically-significant PDEs. Characteristic Problem. Integrability. Brackets algebras. Shape Theory. Foundations of Geometry. 

\vspace{0.1in}
  
\n PACS: 04.20.Cv, 04.20.Fy, Physics keywords: observables, generalized Killing vectors, Background Independence. Problem of Observables facet of Problem of Time. 

\vspace{0.1in}
  
\n $^*$ Dr.E.Anderson.Maths.Physics@protonmail.com

\section{Introduction}

\n Consider a differentiable manifold \cite{Lee2, KN1} 

\n\be 
\bFrM  \m .
\ee 
The corresponding automorphisms are the diffeomorphisms 

\n\be 
\varphi: \bFrM \m \longrightarrow \m \bFrM \m , 
\ee
forming the group  
\be 
Aut(\bFrM) = Diff(\bFrM)     \m .  
\label{Diff}
\ee
The differentiable structure of $\bFrM$ supports \cite{Lee2, KN1} notions of scalar, vector, and more generally tensor fields, 
and other geometrical objects, collectively denoted $\bG$, each subject to a particular transformation law. 
It moreover  supports a notion of {\it Lie derivative} \cite{Stewart, Yano55} with respect to a vector field $\u{X}$, 

\n\be 
\pounds_{\u{X}}          \m . 
\ee 
This generates the local infinitesimal version of the diffeomorphisms -- forming an infinite-dimensional Lie algebra -- and also acts on the $\bG$.    
If one enters the infinitesimal transformation 

\n\be 
\u{x} \m \longrightarrow \m \u{x}^{\prime}  \es  \u{x} + \epsilon \, \u{X}
\ee 
into a particular $\bG$'s transformation law, 
\be 
\pounds_{\u{X}} \bG = 0 
\label{LG}
\ee 
arises to first order \cite{Eisenhart33, Yano55}.  

\m 

\n Some of the tensors and other geometrical objects moreover have further significance as further levels of geometrical structure, $\bsigma$.   
Examples include \cite{Yano70, Yano55, KN1, Kobayashi} the metric tensor and the connection, 
as well as simlarity and conformal equivalence classes of metrics and projective equivalence classes of connections.
Let us use the notation 

\n\be
\langle \bFrM, \bsigma \rangle
\ee
for a differentiable manifold $\bFrM$ equipped with geometrical level of structure $\bsigma$.  
Such cases' version of (\ref{LG}), 

\n\be 
\pounds_{\u{X}} \bsigma = 0  \m , 
\ee 
is known furthermore as a {\it generalized Killing equation} \cite{Yano55, Yano70}.  
The solutions thereof are {\it generalized Killing vectors} $\xi_a = \xi_a(\u{x})$ (for $\xi_a$ the covector corresponding to $X^a$ \cite{Yano70}).
For a particular $\langle \bFrM, \bsigma \rangle$ these moreover close \cite{Kobayashi} as a Lie algebra, 

\end{titlepage}

\n\be
\mbox{\bf [} \xi_A(\u{x}) \mbox{\bf ,} \, \xi_B(\u{x}) \mbox{\bf ]}  \es  {C^C}_{AB} \, \xi_C(\u{x})           \m , 
\label{xi-Lie}
\ee
where $A, B, C$ are multi-indexes comprising both the corresponding spatial index $a, b, c$ 
and the generator-basis index $g$, and ${C^C}_{AB}$ are the corresponding {\it structure constants}.
This Lie algebra corresponds to the continuous part of the automorphism group,  
\be 
Aut(\bFrM, \bsigma)  \m . 
\label{Aut}
\ee 
{\it Automorphism equations} solved by {\it automorphism vectors} provides conceptually useful aliases for the above.  
For later use, we also note that furthermore the infinitesimal-generator form of automorphism vectors
\be
\xi_A  \es  {G_A}^b(\ux)\frac{\pa}{\pa x^b}   \m ,
\ee
and that the (\ref{Aut}) are finite \cite{Eisenhart33, Yano55} subalgebras of the (\ref{Diff}), in the sense of generator count. 

\m 

\n We consider furthermore the notion of {\sl constellations} of $N$ points on a given manifold $\langle \bFrM, \bsigma \rangle$.  
Ab initio, these form product-space {\it constellation spaces}  
\be
\FrQ(\bFrM, \bsigma, N)  \:=  \bigtimes_{I = 1}^N \, \langle \bFrM , \, \bsigma \rangle             \m . 
\ee 
{\sl Shape Theory} \cite{Kendall84, Kendall89, Small, Kendall, Bhatta, DM16, PE16, I, II, III, Minimal-N} considers moreover the effect of quotienting this by (\ref{Aut}), with 
\be 
\FrS(\bFrM, \bsigma, N)  \:=  \frac{ \bigtimes_{I = 1}^N \, \langle \bFrM , \, \bsigma \rangle }{ Aut(\bFrM, \bsigma) }
\ee 
defining the corresponding {\sl shape space}.  
(We denote configuration spaces more generally by $\FrQ$.)  
These notions apply also \cite{LR95, LR97, M02, M05, FileR, M15, Minimal-N} in the physical context, 
in which the $N$ points are materially-realized by $N$ particles: an $N$-Body Problem \cite{Marchal, LR97} (nonrelativistic, and classical in the current Article's context).

\m 

\n More specifically, the current Series of Articles \cite{PE-2, PE-3, PE-4, PE-5} considers preserved quantities by {\sl Taking Function Spaces Thereover}.
Understanding this choice of expression begins with considering the following 2 cases.  

\m 

\n{\bf Case A)} $\langle \bFrM, \bsigma \rangle$ has no nontrivial continuous automorphisms. 
This is moreover the generic situation, as explained below and in Sec 3's outline.    
In this case, 
\be 
\mbox{any function on } \m \bFrM  
\ee 
is a preserved quantity. 
For coordinates $q^a$ valid in some patch on $\bFrM$, this means that any 
\be 
\sbiQ = \sbiQ(q^a)
\ee 
are local preserved 1-point quantities on $\bFrM$, and any 
\be 
\sbiQ = \sbiQ(q^{aI})
\label{P-free}
\ee 
are local preserved $N$-point quantities on the constellation space $\FrQ(\bFrM, N)$. 
We term this procedure {\it Taking a Function Space Thereover}, with reference to over the configuration space, $\FrQ$. 
For the purposes of doing Differential Geometry or Physics, we do require these functions to be suitably smooth (we consider ${\cal C}^\infty$ functions in the current Series).  
So the function spaces in question are 
\be 
{\cal O}(\bFrM, N)  \:=  {\cal C}^{\infty}(\FrQ(\bFrM, N))  \es  {\cal C}^{\infty} \left( \bigtimes_{I = 1}^N \, \bFrM  \right)  \m . 
\ee 
\n{\bf Case B)} $\langle \bFrM, \bsigma \rangle$ possesses nontrivial continuous automorphisms. 
This case is non-generic, but requires a whole lot more work. 
Use of high-symmetry spaces is moreover an almost-ubiquitous modelling assumption in STEM 
(rightly or wrongly, case by case, since some uses entail appropriate modelling, while others are carried out for the ab initio simplicity and familiarity of such workings).  

\m 

\n In Case B), the a priori free functions (\ref{P-free}) are subject to {\it preservation equations}, arising as zero Lie brackets 
\be 
\mbox{\bf[} \scS_{A} \mbox{\bf,} \, \sbiQ \mbox{\bf]}  \es  0  \m ,
\label{SP}
\ee 
with the {\it sum-over-points} of each particular generator,  

\n\be 
\scS_A   \:=  \sum_{I = 1}^N  {G_A}^b(q^{Ic}) \frac{\pa}{\pa q^{b I}} \m .  
\label{S-Def}
\ee
{\bf Theorem 1} Preserved equations moreover take the form of a {\sl system} of equations, 
with only the generators corresponding to consistent automorphism (sub)algebra forming a valid such system. 

\m 

\n{\u{Derivation}} This subalgebra criterion follows from Jacobi's identity acting on one copy of $\sbiQ$ and two of $\scS_A$: 
\be 
             \mbox{\bf{[}} \,\mbox{\bf{[}}  \scS_A \mbox{\bf ,} \, \scS_B \mbox{\bf],} \, \sbiQ \mbox{\bf ]}   \es  
- \left\{ \, \mbox{\bf{[}} \,\mbox{\bf{[}}  \scS_B \mbox{\bf ,} \, \sbiQ   \mbox{\bf ],} \, \scS_A \mbox{\bf ]} \m + \m 
             \mbox{\bf{[}} \,\mbox{\bf{[}}  \sbiQ  \mbox{\bf ,} \, \scS_A \mbox{\bf ],} \, \scS_B \mbox{\bf ]}  \,  \right\} 
																			                                          \es  0 \m + \m  0 
																			                                          \es   0                                    \m ,
\ee 
so $\sbiQ$ must also form 0 Lie brackets with $\mbox{\bf[} \scS_A \mbox{\bf ,} \, \scS_B \mbox{\bf]}$, so the quantities forming zero brackets brackets-close.   $\Box$   

\m

\n{\bf Corollary 1} Preserved quantities can only be posited for subsets of generators that are known to close. 

\m  

\n Case B) is one of the ways in which `Taking Function Spaces Thereover' passes from 
   Case A)'s apparent triviality to a working that is not only nontrivial but a fortiori {\sl foundational}, as argued below and in \cite{ABook, 8-Pillars, 5-6-7}.  

\m 
 
\n The main point of the current Series is that the Lie bracket equation (\ref{SP}) can furthermore be written out as an explicit PDE system (Sec 3), 
which moreover has many tractable features and subcases  that we subsequently exploit.  

\m 

\n This PDE is moreover a subcase of that for canonical observables in Theoretical Physics. 
We explain this point further in Sec 2 with examples of notions of observables and the corresponding PDEs, 
as well as arguments that the current Article's PDE classification and methodology transcends to much of this case as well. 
In a nutshell, 

\m 

\n 1) These quantities all arise by Taking Function Spaces Thereover. 

\m 

\n 2) They all solve zero brackets equations in the presence of first-class constraints $\sbcF$ or sums-over-points $\sbcS$. 

\m 

\n 3) These brackets equations can moreover be written out as explicit first-order linear PDE systems.

\m 

\n 4) In the secondmost-generic case of a single $\scF$ or $\scS$, there is just a single first-order linear PDE. 
Standard `flow' approaches are available to reduce this single PDE to a system of first-order ODEs, including admission of a standard treatment for Characteristic Problems.

\m 

\n 5) We furthermore provide firm grounding for free characteristic treatment's appropriateness, regardless of whether one has just the one PDE or a system.  
`Free', alias `natural' \cite{CH1}, signifies specifically treating the general problem rather than subjecting it to prescribed data. 
This procedure is, on the one hand, geometrically and physically appropriate. 
On the other hand, it embodies Taking Function Spaces Thereover in 2)'s context. 
 
\m 

\n 6) In the case of a nontrivial system -- $\geq 2$ equations -- over-determinedness occurs, since it is still to be solved for a single function. 
Determinedness is indeed one reason why, unlike for single PDE, there is not a universally applicable method for nontrivial systems of PDEs. 
Over-determined cases must meet integrability conditions if they are to possess proper solutions (as opposed to any homogeneously-guaranteed trivial solutions). 
For example, automorphism equations alias generalized Killing equations are also over-determined PDE systems, 
and these almost never meet the corresponding integrability conditions \cite{Yano55}. 
This is indeed why the generic manifold possesses no (generalized) Killing vectors.  

\m 

\n 7) For preserved equations, however, we prove in Sec 3 that {\sl suitable integrability conditions are always met}.   
This can be summarized by the generic manifold, having {\sl no} generalized Killing vectors but {\sl free} preserved quantities.
We also show that, in the non-generic case, observables equations moreover follow suit as regards integrability. 
This furnishes a {\sl general Local Existence Theorem} for geometrical preserved quantities -- and classical-physics finite-model canonical observables -- 
provided that the point-or-particle number $N$ is large enough (a matter of needing {\sl some} local coordinates to make nontrivial, i.e.\ nonconstant, functions out of). 

\m

\n 8) We also show that (Sec 4) preserved quantities -- like observables \cite{ABeables} -- themselves form an algebraic structure.

\m 

\n 9) Sec 5 outlines how the preserved-quantity algebraic structure's subalgebraic structures are moreover in 1 : 1 correspondence with those of the $\scS_A$ and thus of the $\xi_A$. 
They form a bounded lattice of preserved-quantity subalgebraic structures dual to that of $\scS_A$ and thus of the $\xi_A$, i.e.\ as formed by the automorphisms.
Similarly, observables algebraic structures form a bounded lattice dual to that of constraint algebraic structures.  

\m 

\n 10) The Free Characteristic Problem posed in 5) leads to consideration of intersections of characteristic surfaces, 
which can moreover be conceived of in terms of restriction maps. 
This clarifies that Taking Function Spaces Thereover in fact carries {\sl Presheaf-Theoretic} \cite{Wells, Sheaves1, Sheaves2} connotations in a natural manner (Sec 6). 
For presheaves are spaces over a space that are inter-related by restriction maps, and the situation in hand already comes equipped with manifest and meaningful restriction maps. 
On the one hand, presheaves have already been used to formulate quantum observables \cite{FH87}, and other quantum-level applications \cite{ID-Topos}.
On the other hand, there may be some chance of promotion to a Sheaf-Theoretic formulation 
as regards providing a global counterpart to the current Article's local classical treatment of preserved quantities and observables.    
It is also worth commenting that it is becoming quite standard \cite{Wells, Ghrist} in advanced considerations of function spaces to have multiple function spaces over a given  
underlying  `base space', and to model this situation using (pre)sheaves. 
`Taking Function Spaces Thereover' is thus an expression designed to carry the implications of, firstly,  associating a function space to an underlying base space, and, secondly, 
of doing so multiple times (for subspaces of the base space).
In situations which come with natural restriction maps between these function spaces as well, one can furthermore entertain `{\it Taking a Presheaf of Function Spaces Thereover}'. 
If suitable global-gluing, and localization, are furthermore associable, one would be a fortiori modelling by `{\it Taking a Sheaf of Function Spaces Thereover}'; 
this however lies outside the scope of the current Series.    
%
{            \begin{figure}[!ht]
\centering
\includegraphics[width=0.6\textwidth]{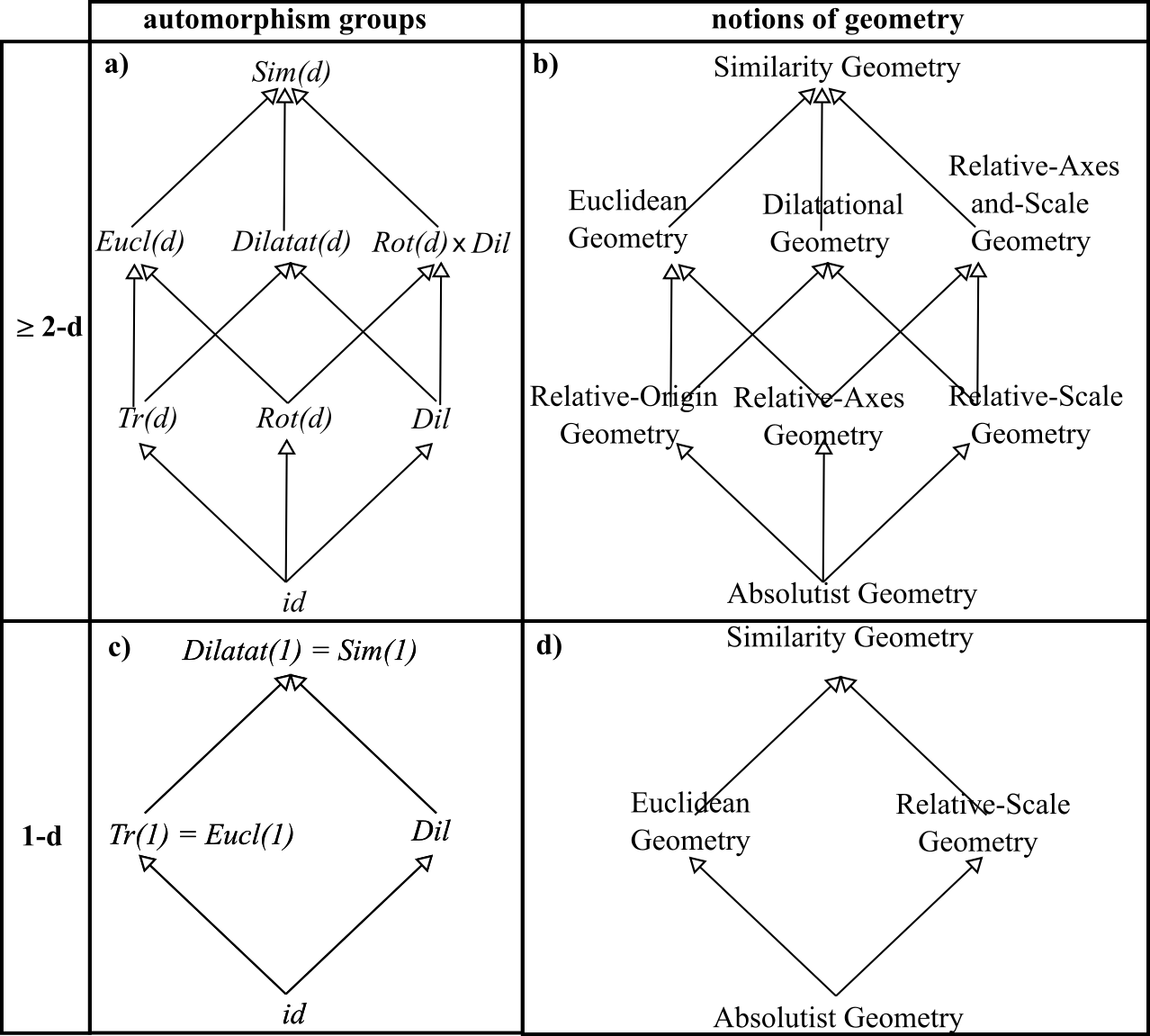}
\caption[Text der im Bilderverzeichnis auftaucht]{        \footnotesize{Lattices of a) this Article's geometrical automorphism groups and b) the corresponding geometries. }}
\label{Sim-Latt} \end{figure}          }

\m 

\n Secs 7 to 15 provide some of the simplest examples of actually solving specific preserved equation (systems). 
We here build up the preserved quantities for Similarity Geometry via the similarity group $Sim(d)$'s geometrically-significant subgroups.
Namely, just translations $Tr(d)$, 
        just dilations $Dil$,  
	and just rotations $Rot(d)$, 
alongside the translations-and-dilations of dilatational geometry, 
the translations-and-rotations of Euclidean geometry,
and rotations-and-dilations geometry, noted for its two types of transformations acting independently.  
These examples form the bounded lattice of automorphism subgroups of Fig 1.a), with Fig 1.b) providing the corresponding lattice of geometries; 
Figs 1.c-d) illustrate the corresponding simplifications following from working in 1-$d$. 
These examples moreover already serve to illustrate, firstly, solving single preserved equations by the standard method available to these.  
Secondly, `blockwise' and `sequential' methods in sufficiently decoupled systems, 
alongside decoupling changes of variables by which `factorizability into blocks' is on occasion attained.  
Thirdly, `compatibility equations' seeking to combine individual equations' functional-form solutions into joint functional-form solutions. 
Fourthly, `chain rule sequential methods', 
by which one substitutes a coordinate change into `coordinates that include the characteristics of a subset of the equations' into the remaining equations.

\m 

\n The Conclusion (Sec 16) then extends Fig 1 to include the dual bounded lattice of notions of preserved quantities, 
alongside the corresponding presheaves thereover (we take the 1-$d$ case to suffice to illustrate such presheaves).  
It finally points to many further applications of the current work, including to 
the Foundations of Geometry                        \cite{Hilb-Ax, HC32, Coxeter, S04, Stillwell}, 
and to observables                                 \cite{DiracObs, Kuchar92, I93, Kuchar93, ABeables, AObs2, AObs3, ABook} 
and Background Independence in Theoretical Physics \cite{A64, Dirac, A67, Battelle, DeWitt67, BB82, HT92, Kuchar92, I93, Giu06, APoT2, APoT3, ABook, PoT-Lett}.

\vspace{10in}

\section{On preserved quantities' extension to observables}

Fig \ref{PE-OE} provides the `big picture' and notation for this.
Some further features therein require the following explanations. 
%
{            \begin{figure}[!ht]
\centering
\includegraphics[width=0.97\textwidth]{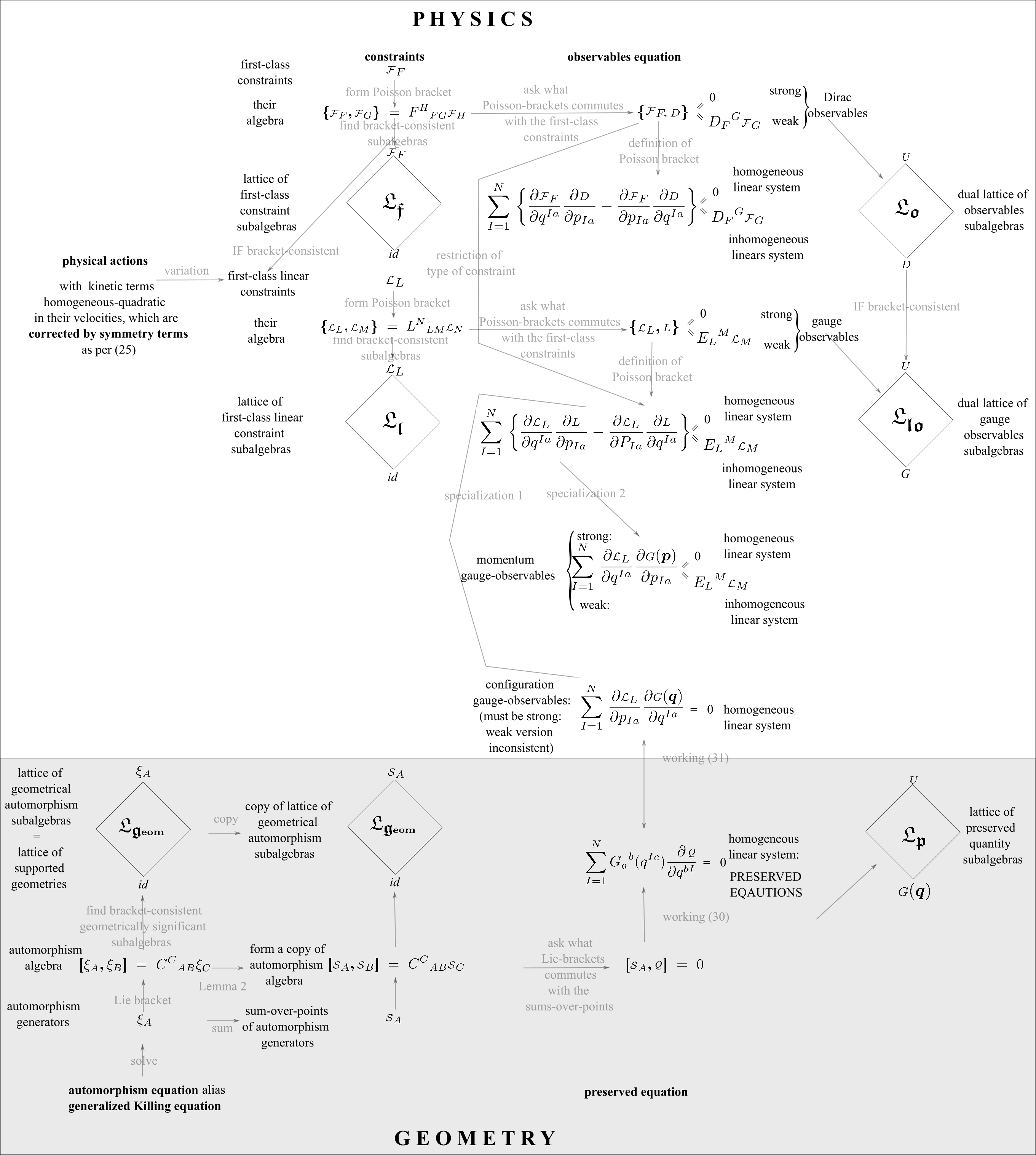}
\caption[Text der im Bilderverzeichnis auftaucht]{        \footnotesize{Conceptual scheme for how preserved equations and observables equations arise, 
with inter-relations indicated in grey. 
Constraints are collectively picked out as a conceptual class of objects by being denoted by undersized calligraphic letters: 
$\tbcF$ for first-class, and $\tbcL$ for first-class linear; sums-over-points of generators also belonging to this conceptual class, these are denoted by $\tbcS$. 
Observables are collectively picked out as a conceptual class of objects by being denoted by undersized italic letters: 
$\tbiO$, $\tbiU$, $\tbiD$ and $\tbiG$ for general, unrestricted, Dirac and gauge observables respectively. 
Geometrical preserved quantities also belonging to this conceptual class, these are denoted by $\tbiQ$. 
This figure `paints the big picture' of where the current Article's preserved equations come from 
and that canonical observables equations provide technically closely related observables PDEs. 
A subset of it moreover also sets up Theorem 2 that Physics' configuration gauge-observables are one and the same as Geometry's preserved quantities: $\tbiG(\sbiq) = \tbiQ$. 
This is the means by which the geometrical study of preserved quantities extends to the physical study of canonical observables, 
alongside the justification of why the $\tbiQ$ belong to the same conceptual class as the $\tbiO$. }}
\label{PE-OE} \end{figure}          }

\m

\n{\bf Remark 1} On the one hand, `weak' is meant in the sense of Dirac \cite{Dirac}, meaning zero `up to a homogeneous-linear function in the constraints'. 
This provides weak observables PDEs' inhomogeneous term.
On the other hand, `strong' just means identically zero.  

\m

\n{\bf Remark 2} The configurational case cannot however support any properly weak observables. 
This is because we are in a context in which constraints have to depend on momenta
\be 
\scC = f(\biq, \, \bip) \m \mbox{ \sl with specific $\bip$ dependence } \m ;  
\ee
a fortiori, first-class linear constraints refer specifically to being linear in their momenta $\bip$, 
By this, the weak configurational observables equation would have a momentum-dependent inhomogenous term right-hand side. 
But this is inconsistent with admitting a solution with purely $\biq$-dependent right-hand-side. 

\m 

\n{\bf Side-Caveat} While first-class linear constraints and gauge constraints are not always exactly the same, 
by which Kucha\v{r} observables $\sbiK$ \cite{Kuchar93} commuting with the former can be distinct from gauge observables $\sbiG$ commuting with the latter, 
this distinction does not enter the current Series' geometrical examples, and so is left to other more physical accounts \cite{HT92, ABook}.

\m 

\n{\bf Remark 3} Let us next clarify that strong configuration observables corresponding 
to first-class linear constraint algebras that constitute geometrically significant automorphism groups {\sl are} geometrical preserved quantities. 

\m 

\n{\bf Theorem 2}  Physical configuration gauge-observables coincide with geometrical preserved quantities, 
\be 
\sbiG(\biq) = \sbiQ  \m . 
\ee
Our proof of this proceeds via a Lemma and a Corollary, as follows. 

\m 

\n{\bf Lemma 1} Let $S$ be a finite action whose kinetic term is homogeneous-quadratic in its velocities $\dot{\u{q}}^I$.
Let its velocities be furthermore corrected by generators ${G_A}^b$ of some geometrical automorphism group $\lFrg$ acting on $\u{q}^I$ multiplied by auxiliaries $\alpha_{A}$.  
Then the ensuing constraints from variation with respect to the $\alpha_{A}$ are 

\n\be 
\scL_A   \es  \sum_{I = 1}^N {G_A}^b(\u{q}^I) p_{bI}   \m .  
\ee
These are in particular linear in their velocities and first-class.  
%

\m 

\n{\u{Derivation}} 
\be 
S  \es  \int L(\biq, \dot{\biq}) \, \d t  
   \es  \int \left( \, \frac{1}{2}M_{ab}\dot{q}^{aI}_{\alpha}\dot{q}^{bI}_{\alpha} - V(\biq) \right) \d t
\ee
for Lagrangian $L$, homogeneous-quadratic kinetic metric $M_{ab}$ and potential $V$. 
The 

\n\be
\dot{q}^{aI}_{\alpha}  \:=  \dot{q}^{aI} - \sum_{g \, \, \in \, \, \nFrg} {G^a}_{b g} \alpha^{b g}
\ee 
are furthermore the alluded-to corrected velocities.  
The conjugate momenta are 
\be 
p_{aI}  \:=  \frac{\pa L}{\pa \dot{q}^{aI}} 
        \es   M_{ac} \, {\dot{q}^c}_{I\alpha}     \m . 
		\label{mom}
\ee
The outcome of varying our action with respect to $\alpha^{a g}$ is 

\n\be
0 = \scC_{a g}  \:=  \frac{\pa L}{\pa \alpha^{a g}} 
                \es  \sum_{I = 1}^N {G_a}^{b g}(\uq^{I})M_{bc}\dot{q}^{cI}_{\alpha} 
		        \es  \sum_{I = 1}^N {G_a}^{b g}(\uq^{I})p_{bI}
\ee 
by multiplier equation simplification, then homogeneous quadraticity of the action, and then (\ref{mom}). 
Finally introducing the multi-indices $A := a \, g$ etc.\ yields the desired answer. 
This is manifestly linear, and closes with itself under Poisson brackets \cite{Dirac, Sni} as required, yielding a representation of the underlying Lie algebra,  
\be 
\mbox{\bf \{} \scL_A \mbox{\bf ,} \,  \scL_B \mbox{\bf \}}  \es  {L^C}_{AB} \, \scL_C   \m , 
\ee
thus establishing first-classness, as required. $\Box$

\m 

\n{\bf Corollary 1} 
\be 
\pa_{P_{bI}} \scC_A = {G_A}^b(\u{q}^I) \m . 
\ee
\n{\u{Proof of Theorem 2}} On the one hand, the $\sbiQ$ are the suitably-smooth functions obeying the preserved equation 

\n\be 
0  \es  \mbox{\bf [}\scS_a^g  \mbox{\bf ,} \, \sbiQ  \mbox{\bf ]}  
   \es  \sum_{I = 1}^N {G_A}^{b}(q^{Ic}) \pa_{q^{Ib}} \sbiQ                 \m . 
\ee 
On the other hand, the $\sbiG(\biq)$ are the likewise suitably-smooth functions obeying the configuration gauge-observables equation 

\n\be 
0  \es    \mbox{\bf \{} \scL_A  \mbox{\bf ,} \, \sbiG(\biq)  \mbox{\bf \}} 
   \es  \sum_{I = 1}^N \pa_{p_{Ia}} \scL_A(q^{Jb}) \, \pa_{q^{Ia}} \sbiG(\biq)
   \es  \sum_{I = 1}^N {G_A}^{b}(q^{Ic}) \, \pa_{q^{Ia}}           \sbiG(\biq)   \m , 
\ee 
where the second equality is by evaluating the Poisson bracket with $\pa_{p_{Ia}}$ annihilating $\sbiG = \sbiG(\biq \mbox{ alone})$, and the third follows from the Corollary.
Thus the $\sbiQ$ and the $\sbiG(\biq)$ form the same function space, and so are an identical notion. $\Box$ 

\m 

\n{\bf Remark 4} This is the `Bridge Theorem' between preserved quantities and observables alluded to in the Introduction.  

\m 

\n{\bf Remark 5} Theorem 1 moreover admits a sweeping generalization across this bridge.  

\m 

\n{\bf Theorem 1$^{\prime}$} Observables equations -- canonical or spacetime, classical or quantum -- moreover take the form of a {\sl system} of equations, 
with only the generators corresponding to consistent automorphism (sub)algebras forming a valid such system. 

\m 

\n{\bf Remark 6} Spacetime observables are ones that commute with spacetime generators under the Lie bracket operation that the latter close under.

\m 

\n{\bf Remark 7} In the above context, Theorem 1$^{\prime}$ is the `C C B Theorem', standing for putting 2 constraints and 1 beable into the Jacobi identity. 
(My concept of `beable' coincides at the classical level, with that of observable.) 
This Theorem first appeared in \cite{ABeables}.
\cite{ABook} contains the spacetime observables counterpart of this result, and terms it `the first great decoupling' in providing a local resolution of the Problem of Time.  
We analogously call Theorem 1 itself the `S S Q Theorem'.

\section{Characteristic Problems for first-order PDEs}

\n{\bf Structure 1} The current Series' preserved equation systems are in general of the form g) in Fig 3, 
for {\it one independent variable} $\phi = \sbiQ$ in terms of $\alpha = 1$ to $k$ {\it dependent variables} $x^{\alpha}$.   
For our particular problem, we use $\sbiQ$ for $\phi$, standing for `preserved quantity', and the role of the $k$      quantities $x^{\alpha}$ 
                                                                                            is played by the $N \, d$ quantities $q ^{Ia}$.  
%
{            \begin{figure}[!ht]
\centering
\includegraphics[width=0.6\textwidth]{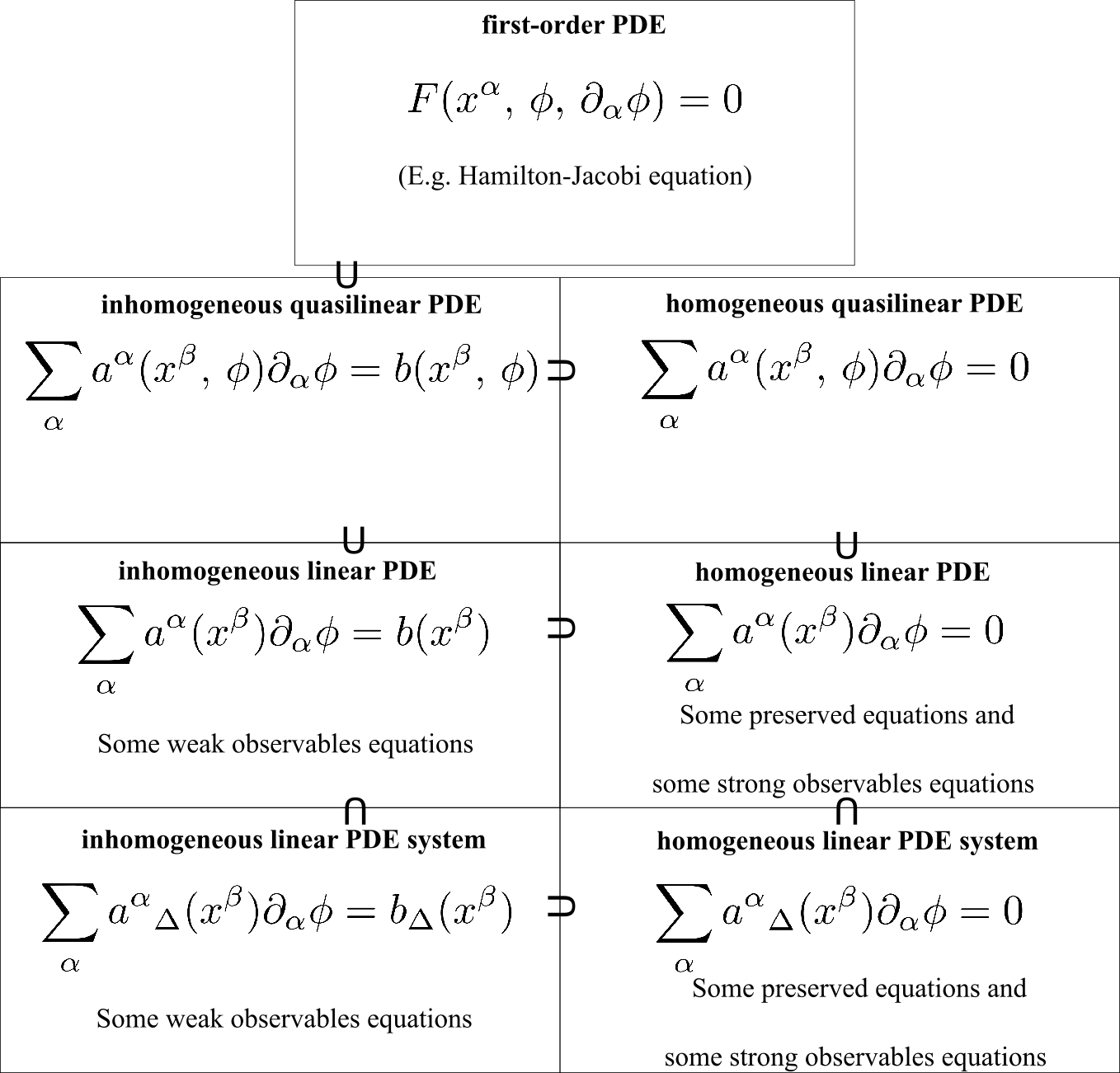}
\caption[Text der im Bilderverzeichnis auftaucht]{        \footnotesize{a) first-order PDEs. 
                                                                        b) First-order inhomogeneous quasilinear PDEs. 
																		c) First-order homogeneous quasilinear PDEs. 
																		d) First-order inhomogeneous linear PDEs.
                                                                        e) First-order homogeneous linear PDEs.
 																		f) First-order inhomogeneous linear PDE systems.
                                                                        g) First-order homogeneous quasilinear PDE systems.
                                                                        In each case, geometrical or physical examples are given. 
Fig 2 derives all but the parenthesized	examples of these equations indicated in the current figure.  																
 																		}}
\label{Leib(3, 2)-Full-Jac} \end{figure}          }

\m  

\n{\bf Structure 2} Forms a) to f) of Fig 3 are moreover useful toward classifying and solving these \cite{CH2, John, Lee2}, 
as well as their counterparts from Physics, the canonical observables equations. 
All of these equations lie within the class of {\it first-order} PDEs: form a). 
This refers to their containing first-order partial derivatives $\pa_{\alpha}\phi$ and no higher.  

\m 

\n{\bf Structure 3}  A further specialization of note -- since much PDE theory has been developed for it -- is to the first-order {\it quasilinear} equations: form b) of Fig 3.
This refers to the PDE being linear in the $\pa_{\alpha}\phi$, for all that it can still be nonlinear in $\phi$ itself.  

\m  

\n{\bf Structure 4} Among these, the general case is the {\it inhomogeneous} equation of Fig 3.b) itself, whereas its {\it homogeneous} specialization if of the form c) in Fig 3.  
This refers to homogeneous linearity in the $\pa_{\alpha}\phi$. 
One consequence of this is that the constant function,  
\be
\phi = \mbox{const}  \m ,
\ee
always solves; we refer to this as the {\it trivial solution}, and to all other solutions of first-order homogeneous quasilinear PDEs as {\it proper solutions}. 
Such are supported if the corresponding differential operator has {\it nontrivial kernel}.  

\m  

\n{\bf Structure 5} A further specialization is to first-order (in)homogeneous {\it linear} PDEs -- inhomogeneous form d) and homogeneous form e) in Fig 3 -- 
referring to full linearity in $\phi$, i.e.\ linearity in both $\pa_{\alpha}\phi$ and in $\phi$ itself in this first-order case.

\m 

\n One of the simplest cases of preserved equations is when they are of form e), whereas two of the simplest cases of observables equations are of forms d) and e). 
Observables equations that are not themselves preserved equations that are of form e) include single strong momentum observables equations 
                                                                                          and single strong general observables equations.
Observables equations of form d) include single {\sl weak} configuration, momentum and general observables equations.   

\m 

\n{\bf Remark 1} There is a well-established theory of PDEs for all of forms a) to e), which is increasingly effective as one goes down and right along Fig 3's grid 
due to the further simplifying specializations. 
%
{            \begin{figure}[!ht]
\centering
\includegraphics[width=0.6\textwidth]{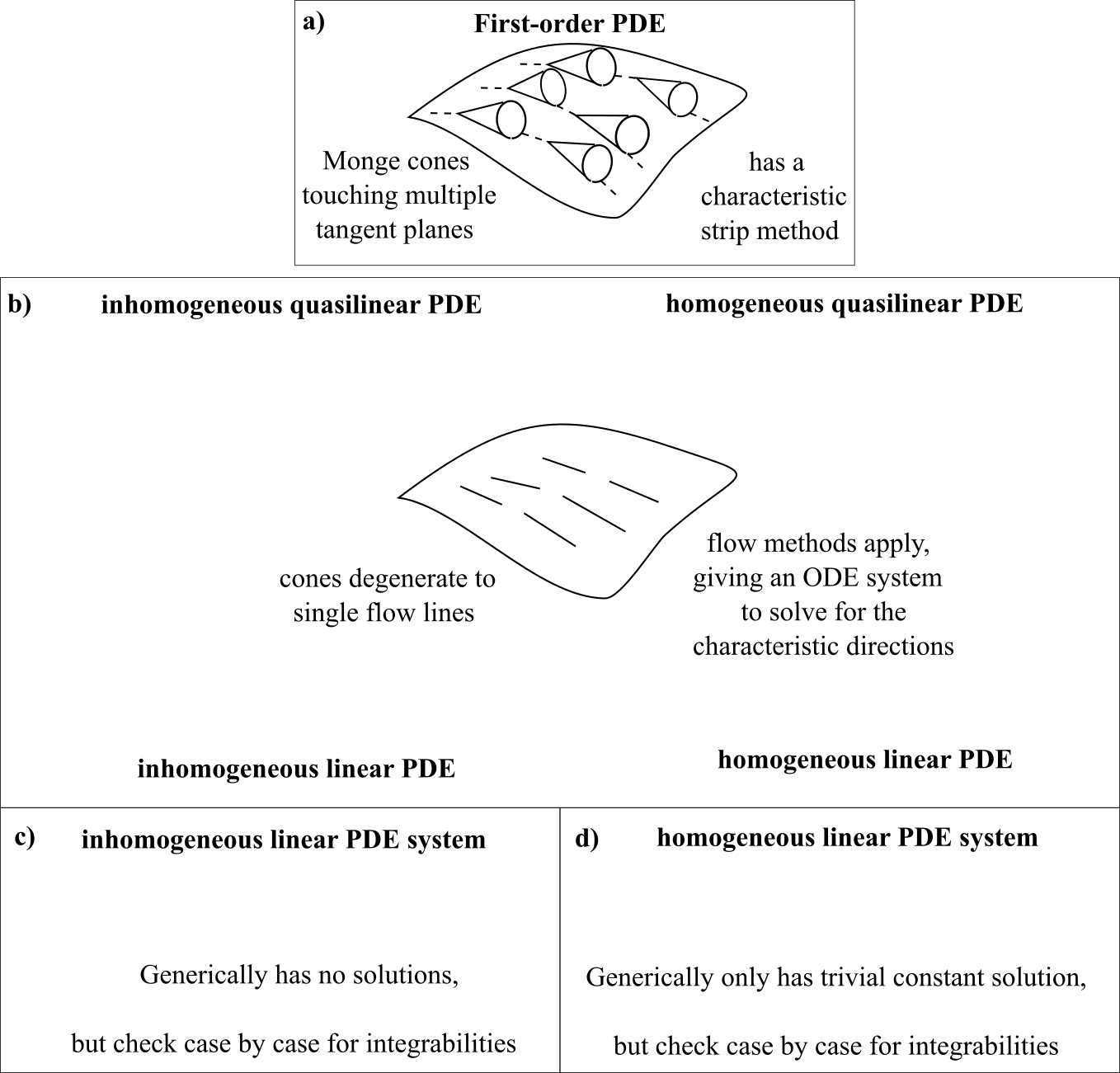}
\caption[Text der im Bilderverzeichnis auftaucht]{        \footnotesize{Characteristic treatments for each of Fig 3's PDEs \cite{John, CH2}.}}
\label{3-Box} \end{figure}          }

\m 

\n{\bf Remark 2} In the case of {\it Characteristic Problems} \cite{CH2, John}, as occur in the current Series, how to proceed is indicated in Figs 4.b-d).  
The {\it free} alias {\it natural} \cite{CH1} version of this involves finding the general solution rather than imposing specific prescribed data. 
This returns the full diversity of preserved quantities supported by the system. 

\m 

\n This is appropriate since we are Taking a Function Space Thereover -- meaning over a geometry, a configuration space -- 
more specifically a reduced configuration space: shape space -- a phase space, again more specifically a reduced phase space: shape phase space.
This involves finding {\sl all} such functions, say withing a given smoothness category such as ${\cal C}^{\infty}$, rather than a particular such corresponding to prescribed data. 

\m 

\n{\bf Remark 3} What we want for now, more specifically, is how to deal with Characteristic Problems for equations of form 2) in Figure 3. 
The current Article just considers a local rather than global treatment of preserved equations.
It turns out that this method approach immediately generalizes as far back as form b) of Fig 2: inhomogeneous quasilinear equations, so we present this case at no extra cost. 
The corresponding characteristics of this are determined by 
\be 
\frac{\d x_{1}}{a_{1}(x^{\beta}, \phi)}  \es  \frac{\d x_{2}}{a_{2}(x^{\beta}, \phi)}  \es  . \, . \, . \es  \frac{\d x_{k}}{a_{k}(x^{\beta}, \phi)}  \m . 
\ee 
For $k = 2$, this gives an ODE 
\be 
\frac{\d x^1}{\d x^2}  \es  \frac{  a_{1}(x^1, \, x^2, \, \phi)  }{  a_{2}(x^1, \, x^2, \, \phi)  }                                                             \m .
\ee 
In the general case, moreover, our PDE is equivalent to the ODE system 
\be 
\dot{x}^{\alpha} = a^{\alpha}(x^{\beta}, \phi)                                                                                                        \m , 
\ee 
\be 
\dot{\phi} = b(x^{\beta}, \phi)                                                                                                                       \m .
\ee
Here 
\be
\dot{\m}   \:=  \frac{\d}{\d t}                                                                                                                       \m , 
\ee
for $t$ a fiducial variable to be eliminated, rather than carrying any temporal (or other geometrical or physical) significance; 
our system is moreover autonomous (none of the functions therein depend on $t$).  
This linear ODE system has well-understood local existence properties. 

\m 

\n{\bf Remark 4} In the homogeneous-linear case, our system simplifies further to 
\be 
\dot{x}^{\alpha} = a^{\alpha}(x^{\beta})   \m , 
\ee  
\be 
\dot{\phi} = 0                             \m . 
\ee 
This last of these equations is solved by 
\be 
\phi = \phi(u^{\bar{\alpha}})              \m , 
\label{phi(u)}
\ee 
where barring an index denotes it taking one value less than the unbarred version, and the $m - 1$ quantities $u^{\bar{\alpha}}$ are a basis of characteristic coordinates.  

\m 

\n The integrated form of the first $m$ equations is used to eliminate $t$, 
with the other $m - 1$ each providing a characteristic coordinate arising as its constant of integration. 
One then substitutes these characteristic coordinates into (\ref{phi(u)}) to obtain the general -- and thus free alias natural problem-solving -- characteristic solution.  

\m  

\n{\bf Structure 6} There is moreover a sense of middling genericity in which preserved equations, and observables equations, are first-order linear {\sl systems}, 
whether homogeneous (Fig 2.7) for preserved equations and other strong observables equations, or inhomogeneous (Fig 2.6) for weak observables.

\m 

\n There is however {\sl no} general treatment for Characteristic Problems for PDE systems of this general kind.  
Further details of the PDE system in question need to be considered in order to proceed, as follows.   

\m 

\n{\bf Structure 7} Preserved equation and observables equation systems consist of of $G = \mbox{dim}(Aut(\bFrM, \, \bsigma))$ equations for a single unknown.
So the default prima facie position is generically one of over-determination \cite{CH2}, leading to no solutions (or only the trivial solution, when guaranteed by homogeneity)
A similar situation arises for generalized Killing equations the number of equations exceeds the number of unknowns there as well. 

\m 

\n{\bf Structure 8} The way such a lack of (nontrivial) solutions might occur is via integrabilities.\footnote{While there is a conceptual counter-acting underdetermination from the characteristicness and the freeness, 
it is our integrability point here that guarantees that things work out.} 
%
On the one hand, generalized Killing equations' integrability conditions \cite{Yano55} are moreover not met generically, 
signifying that there are only any proper generalized Killing vectors at all in a zero-measure subset of $\langle \bFrM, \bsigma\rangle$. 
This corresponds to the generic manifold admitting no (generalized) symmetries.
On the other hand, preserved equations moreover always succeed in meeting integrability, by the following Theorem. 

\m 

\n{\bf Theorem 3} Preserved equations are integrable.

\m 

\n{\bf Lemma 2} For $\xi_a^g$ generating a Lie algebra, the corresponding sums-over-points $\scS_A$ do too, and it is moreover a copy of the same Lie algebra: 
\be 
\mbox{\bf [} \scS_A \mbox{\bf ,} \, \scS_B \mbox{\bf ]}  \es  {C^C}_{AB} \scS_C \m . 
\ee 
{\u{Derivation}} 

\n $$ 
\mbox{\bf [} \scS_A \mbox{\bf ,} \, \scS_B \mbox{\bf ]}  \es  \mbox{\Huge [} \sum_{I = 1}^N \xi_A(\u{q}^I) \mbox{\bf ,} \, \sum_{J = 1}^N \xi_B(\u{q}^J)  \mbox{\Huge ]} 
                                                         \es  \sum_{I = 1}^N \sum_{J = 1}^N \mbox{\bf [} \, \xi_A(\u{q}^I) \mbox{\bf ,} \, \xi_B(\u{q}^J) \, \mbox{\bf ]} 
$$

\n\be
                                                         \es  \sum_{I = 1}^N \sum_{J = 1}^N  {C^C}_{AB} \xi_C(\u{q}^I) \delta^{IJ}
												         \es   {C^C}_{AB} \sum_{I = 1}^N  \xi_C(\u{q}^I)
												         \es   {C^C}_{AB} \scS_C
\ee
by (\ref{S-Def}), 
next linearity, 
next basic differentiation, 
next the constancy of the structure constants alongside the Dirac delta collapsing a sum, 
and finally (\ref{S-Def}) again.  $\Box$ 

\m 

\n{\u{Proof of Theorem 3}} 
The generators of a geometrical automorphism group close as the Lie algebra (\ref{xi-Lie}) 
corresponding infinitesimally to the connected component of the identity of the automorphism group itself. 
Thus by Lemma 2, the corresponding sums-over-points $\scS_a^g$ that form our PDE system in hand close under Lie brackets as well.  
But this is enough to guarantee integrability by the below well-known Theorem. $\Box$ 

\m 

\n{\bf Frobenius' Theorem} \cite{AMP, Lee2} A collection $\bFrW$ of subspaces of a tangent space possesses integral submanifolds iff 
\be 
\forall \m \m \u{X} \, , \, \, \u{Y} \m \in \m \bFrW \mma \mbox{\bf [} \u{X}\mbox{\bf ,} \, \u{Y} \mbox{\bf ]} \m \in \m \bFrW   \m .
\ee 
\n {\bf Theorem 3$^{\prime}$} Classical canonical observables equations are integrable. 

\m 

\n{\u{Derivation}} The preceding argument holds just as well with constraints in place of sums-over-points-of-generators and Poisson brackets in place of Lie brackets. $\Box$
%

\m
 
\n{\bf Remark 5} Thus while proper generalized Killing vectors generically do not exist, 
preserved quantities {\sl always} do, at least in the current Series' local sense, and for sufficiently large point number $N$. 
This last caveat is clear from the examples below, and corresponds to zero-dimensional reduced spaces having no coordinates left to support thereover any functions of coordinates.
This Remark extends to canonical observables equations as well \cite{5-6-7, AObs4}, 
the first of these references containing also further comparison between preserved and observables equations on the one hand and generalized Killing equations on the other.  

\m 

\n 8) As the Introduction pointed out, there is moreover a greater generality. 
Generalized Killing vector nonexistence means no sums-over-points to commute with.
In this case, the most primitive element of preserved quantities -- Taking Function Spaces Thereover -- is manifested in a particularly simple form: 
taking the free functions on the configuration space $\FrQ$. 
We denote free functional dependence by the ${\bm{\sim}}$ symbol.  
The commutation aspect of preserved quantities arises only subsequently, by interplay with symmetries' generators. 

\m 

\n{\bf Remark 6} Interplays moreover consititute the lion's share of Foundational and Theoretical Physics' Problem of Time \cite{Kuchar92, I93, ABook}.  
Observables are one of nine conceptually distinct aspects of this (see the Conclusion for an outline). 
The uncoupled manifestation of finding observables is moreover indeed Taking Function Spaces Thereover.
Only upon theories containing constraints -- arising from other Problem of Time facets -- does commutation with these enter the theory of observables. 
The Conclusion also outlines parallels between the Problem of Time and its resolution by Background Independence on the one hand and the Foundations of Geometry on the other.
This is moreover suggestive that interplays will also represent the lion's share of the parallel work to be done in establishing further Foundations of Geometry. 

\m  

\n 9) Within this secondmost-generic case possessing a single generalized Killing vector, form e) of Fig 3 arises. 

\m 

\n 10) The $\geq 2$-compatible generalized Killing vectors case is only the next most typical, it being here that the systematic method above does not apply. 

\m 

\n How is one to address solving a preserved-equation system? 

\m 

\n{\bf Method I} Solve its equations piecemeal and equate the functional compatibilities thus entailed.   
These {\it compatibility equations} can be viewed as intersections of characteristic manifolds.
The case supporting only the constant function alluded to above, represents a sufficiency of intersecting subspaces that only a zero-$d$ point survives.  

\m 

\n This kind of `piecemeal solution' suffices in various of the simpler cases in the current Series. 

\m 

\n{\bf Method II} ({\it sequential chain rule}) 
Suppose we have two equations. 
Solve one for its characteristics $u^{\beta}$. 
Then substitute 
\be 
\sbiQ = \sbiQ(u^{\beta})
\ee 
into the second equation to find which functional restrictions on the first solution's characteristics the second equation enforces. 
This procedure can moreover be applied inductively, and is independent of the choice of ordering in which the restrictions are applied 
by the nature of restrictions corresponding to geometrical intersections.

\section{Preserved-quantity and observables function spaces are algebras}

\n{\bf Remark 1} Preserved quantities, and observables more generally, carry a generator-valued index of their own: $\siQ_P$ and $\siO_O$.   
This does not enter the preceding consideration of over-determination because all the $\siQ_P$ solve the preserved equation regardless 
($\siQ_P$ is a $\scS_A = \scS_{a \, g}$ scalar: a spatial scalar and a scalar with respect to the original automorphism group). 
Similarly all the $\siO_O$ solve the observables equation regardless 
($\siO_O$ is a $\scC_A = \scC_{a \, g}$ scalar: a spatial scalar and a scalar with respect to the original constraint algebra). 
We mention this internal index at this point due to its entering the algebraic theory internal to the preserved quantities and observables themselves.

\m 

\n{\bf Theorem 4} Preserved quantities close under Lie brackets. 

\m 

\n{\u{Derivation}} This algebra criterion follows from Jacobi's identity acting on two copies of $\sbiQ$ and one of $\sbcS$: 
\be 
\mbox{\bf{[}} \,\mbox{\bf{[}} \siQ_P \mbox{\bf ,}  \, \siQ_Q \mbox{\bf],}  \, \scS_A \mbox{\bf ]}   \es - \left\{ \,  
\mbox{\bf{[}} \,\mbox{\bf{[}} \siQ_Q \mbox{\bf ,}  \, \scS_A \mbox{\bf ],} \, \siQ_P \mbox{\bf ]}   \m + \m  
\mbox{\bf{[}} \,\mbox{\bf{[}} \scS_A \mbox{\bf ,}  \, \siQ_P \mbox{\bf ],} \, \siQ_Q \mbox{\bf ]}     \, \right\} 
                                                                                                    \es  0  \m +  \m 0 
																								    \es  0               \m . 
\ee 
So $\scS_A$ must also form zero Lie bracket with $\mbox{\bf[}\siQ_P\mbox{\bf,}\, \siQ_{Q}\mbox{\bf]}$, 
by which the latter itself obeys the defining property of the $\sbiQ$. 
I.e.\ the $\sbiQ$ are algebraically closed under Lie brackets. $\Box$   

\m 

\n{\bf Remark 2} We write this as 
\be 
\mbox{\bf[} \siQ_P \mbox{\bf ,} \, \siQ_{Q} \mbox{\bf]}  \es  {Q^R}_{PQ} \, \siQ_{R}
\ee 
for {\it preserved quantities algebra structure constants} ${Q^R}_{PQ}$. 

\m 

\n{\bf Remark 3} Theorem 4 also admits a sweeping generalization to observables, as follows.

\m 

\n{\bf Theorem 4$^{\prime}$} Observables -- canonical or spacetime, classical or quantum -- close under Lie brackets. 

\m 

\n At least for the current Series' finite models, this takes the form of a Lie algebra 
\be 
\mbox{\bf[} \siO_O \mbox{\bf ,} \, \siO_{P} \mbox{\bf ]} = {O^Q}_{OP} \, \siO_{Q}  \m , 
\ee
for {\it observables algebra structure constants} ${O^Q}_{OP}$.  

\m 

\n{\bf Remark 4} In the above context, Theorem 4$^{\prime}$ is the `B B C Theorem', 
standing for putting 2 beables and 1 constraint into the Jacobi identity. 
The `B B C Theorem' first appeared in \cite{ABeables}; \cite{ABook} then gave the spacetime observables counterpart of this result;  
we analogously refer to Theorem 4 itself as the `Q Q S Theorem'.  

\m 

\n{\bf Remark 5} Field Theories -- notably ones which are additionally gravitational and/or Background Independent, such as GR and Supergravity -- 
support constraint Lie {\sl algebroids} \cite{Bojowald1, Bojowald2} rather than Lie algebras. 
This accounts for some `algebraic structures' phrasings in the current Article, this term being taken to be a portemanteau of Lie algebra and Lie algebroid cases.  
While brackets-level Theorems 1 and 4 have Field Theory counterparts, 
Sec 3's PDE analysis is moreover severely disrupted by the passage from PDEs to Field Theory's {\sl functional} differential equations (FDEs); 
this makes for a good mid-term focus for follow-up papers extending the current Series' research.

\section{Preserved-quantity and observables {\it{\bf notions}} form bounded lattices}

\n{\bf Theorem 5}  i) Automorphism-generator subalgebras are in 1 : 1 correspondence with preserved quantities subalgebras. 

\m 

\n  ii) Automorphism-generator subalgebras form a bounded lattice \cite{Lattice}, with intersection and span of consistent combinations of generators as meet and join, 
                                                                   $id$ as bottom, 
                                                               and the full automorphism algebra $\scG$ of the $\xi_A$ as top.  
 
\m  
 
\n iii) Preserved quantities subalgebraic structures form the dual bounded lattice, with intersection and joint span of characteristic surface generators as meet and join 
                                                                   the free functions $\siF$ as top = dual-bottom 
															   and the full geometry's notion of preserved quantities $\sbiQ$ as bottom = dual-top. 
															   
\m 

\n{\bf Remark 1}  These are lattices of {\sl notions}, rather than of algebra elements or functions themselves.  
See Fig 1 for examples of ii) and Fig 5 for the corresponding duals exemplifying iii) as well.  

\m 

\n{\bf Remark 2}  One has subalgebraic structures in the context of having a `top group' of automorphisms to work within. 
The similarity group $Sim(d)$ plays this role in the current Article.  
As we shall see in Articles II to V, there are subsequently moreover competing `furtherly top' groups that one can work within.

\m 

\n Passing to observables and with classical-or-quantum generality, we obtain the following counterpart of the above Theorem. 

\m  

\n{\bf Theorem 5$^{\prime}$} i) First-class constraint subalgebraic structures are in 1 : 1 correspondence with canonical observables subalgebraic structures. 

\m 

\n  ii) First-class constraint subalgebraic structures form a bounded lattice, with intersection and span of consistent combinations of generators as meet and join, 
                                                                   $id$ as bottom, 
                                                               and the full first class-algebra of the $\sbcF$ as top.  
 
\m  
 
\n iii) Canonical observables subalgebraic structures form the dual bounded lattice, with intersection and joint span of characteristic surface generators as meet and join 
                                                                the unrestricted observables $\sbiU$ as dual-bottom 
															and the Dirac observables $\sbiD$ as dual-top.
															   
\m 															   

\n{\bf Remark 3} These are likewise lattices of notions of observables.  
	
\m 
														
\n{\bf Remark 4} For some physical theories, the first-class linear constraints close algebraically, 
by which these support the corresponding notion of Kucha\v{r} observables \cite{Kuchar93}.  
 
\m  
 
\n The lattice of notions of observables is moreover a theory-independent generalization of the possibility of there being some kinds of `middling' observables, 
a role played in GR by the Kucha\v{r} observables themselves. 
 
\m  
 
\n{\bf Theorem 5$^{\prime\prime}$} Suppose a given theory's first-class linear constraints form a consistent subalgebra of the first-class constraint subalgebra. 
Then

\m 

\n i) First-class linear constraint subalgebras are in 1 : 1 correspondence with canonical Kucha\v{r} observables subalgebras. 

\m 

\n ii) First-class linear constraint subalgebras form a bounded lattice, with intersection and span of consistent combinations of generators as meet and join, 
                                                                   $id$ as bottom, 
                                                               and the full algebraic structure of first-class-linear constraints as top.  
 
\m  
 
\n iii) Canonical Kucha\v{r} observables subalgebras form the dual bounded lattice, with intersection and joint span of characteristic surface generators as meet and join 
                                                                   the unrestricted observables $\sbiU$ as dual-bottom  
															   and the Kucha\v{r} observables $\sbiK$ as dual-top.
															   
\m

\n{\bf Remark 5} In the below version, `spacetime generators' \cite{ABook} includes part-spacetime-valued internal indices, 
allowing for the incorporation of spacetime formulation of gauge theories.  

\m 

\n{\bf Theorem 5$^{\prime\prime\prime}$} i) Spacetime generator subalgebras are in 1 : 1 correspondence with spacetime observables subalgebras. 

\m 

\n  ii) Spacetime generator subalgebras form a bounded lattice, with intersection and span of consistent combination of generators as meet and join, 
                                                                   $id$ as bottom, 
                                                               and the full algebra of spacetime generators as top.  
 
\m  
 
\n iii) Spacetime observables subalgebras form the dual bounded lattice,  with intersection and joint span of characteristic surface generators as meet and join
                                                                   the unrestricted spacetime observables as dual-bottom  
															   and the full spacetime observables $\sbiS$ as dual-top.

\section{Presheaves of functions over lattices}

\n{\bf Remark 1} At a more advanced level, this `Taking Function Spaces Thereover' can be conceived of in terms of finding {\sl presheaves} of functions.
Consult e.g.\ \cite{Wells} for sheaves of                                           ${\cal C}^{\infty}$ functions over real    manifolds, 
alongside other straightforward examples such as sheaves of holomorphic functions ${\cal C}^{\omega}$           over complex manifolds. 
See in particular \cite{Ghrist} for basic Applied Mathematics applications of (pre)sheaves, \cite{Nash} 
for physical ones and \cite{FH87} for specific -- if quantum -- observables applications of presheaves specifically; 
an outline of what (pre)sheaves are that suitably complements the current Article can be found in Appendix W of \cite{ABook}.  
We however make little use of this more advanced presheaf conceptualization in the current Series, which just attempts a local treatment. 
Sheaves \cite{Hartshorne, Sheaves1, Sheaves2} are moreover a powerful method as regards furnishing global approaches (and in part for localization: 
passing back from global approaches to local ones).     
Sheaves additionally admit differential-geometric and PDE counterparts \cite{Wells, Wedhorn}.
  
\m   

\n{\bf Structure 1} The previous Section's lattices of {\sl notions} of preserved quantities each corresponds to its own function space, with moreover 
restriction maps interrelating these individual function spaces in the manner of a presheaf of function spaces over said lattice.	
This lattice is moreover dual to a lattice of consistent combinations of geometrical $Aut$ generators, or of levels of geometrical structure, 
each with a function space of elements thereover, with restriction maps therebetween also in the manner of a presheaf. 
\n Method II carries moreover a further indication of applicability of presheaves, 
via formulation of the intersections between successive characteristic surfaces as restriction maps.  

\m 
															   
\n{\bf Structure 2} The above moreover also generalizes to lattices of notions of observables each corresponding to its own function space, 
with moreover restriction maps interrelating these individual function spaces in the manner of a presheaf of function spaces over said lattice.	
This lattice is moreover dual to a lattice of consistent combinations of constraints,   
each with a function space of elements thereover, with restriction maps therebetween also in the manner of a presheaf. 

\m 

\n{\bf Example 1}  See the Conclusion's Figure 6 for a simple example of a presheaf of preserved-quantities function spaces over a bounded lattice, 
and the presheaf of sum-over-points of generators' algebraic structures over the dual bounded lattice.
This figure suffices to furthermore identify both of these as contravariant presheaves: 
their restriction maps' arrows run in opposition to the underlying lattice's ordering arrows. 
  
\vspace{10in}  
 
\section{1-$d$ translations}

On $\mathbb{R}$, the translations form 
\be 
Tr(1) = \mathbb{R}  \m . 
\ee
In this case, there is a single preserved equation,  

\n\be 
\sum_{I = 1}^N \pa_I \sbiQ  \:=  \sum_{I = 1}^N \frac{\pa \sbiQ}{\pa q^I} = 0  \m .  
\label{Tr-PE}
\ee 
Let us also introduce the small-model notation 
\be 
x := q_1 \mma y := q_2 \mma z := q_3          \m .
\ee 
For $N = 1$, (\ref{Tr-PE}) reduces to an ODE,  
\be 
\frac{\d \sbiQ}{\d x} = 0                      \m . 
\label{Tr-PE(1,1)}
\ee 
Thus immediately 
\be 
\sbiQ = const                                  \m , 
\ee 
so just the trivial solution is supported. 

\m 

\n $N = 2$ is minimal for (\ref{Tr-PE}) to be a PDE, 
\be 
( \pa_x + \pa_y ) \sbiQ = 0                                                        \m .  
\label{Tr-PE-2}
\ee 
This is a single PDE, so its characteristics follow from  
\be 
\frac{\d x}{1}  \es  \frac{\d y}{1}                                               \m . 
\ee 
Being in 2 independent variables, a direct integration method is available, yielding 
\be 
\d (x - y) = 0   \m \Rightarrow \m x - y = constant                               \m . 
\ee 
This is not moreover a unique solution, since by the chain-rule 
\be 
( \pa_x + \pa_y ) f(x - y)  =  f^{\prime} ( \pa_x x - \pa_y y ) 
                            =  f^{\prime}(1 - 1) = 0                              \m , 
\ee 
where 
\be 
\mbox{}^{\prime}  :=  \d/\d u \m \mbox{ for } \m 
u := x - y \m \mbox{ the problem in hand's characteristic variable }              \m . 
\ee 
Thus 
\be 
\sbiQ  =  \sbiQ(x - y)
\label{P(x-y)}
\ee 
solves, and our preserved quantities are of the sole {\it difference} ({\it relative separation} of points) supported by the system.  

\m 

\n Alternatively, by the more systematically applicable flow method, (\ref{Tr-PE}) is equivalent to the ODE system
\be 
\dot{x} = 1      \m , 
\ee 
\be 
\dot{y} = 1      \m , 
\ee
\be 
\dot{\sbiQ} = 0   \m , 
\ee
to be solved as a Free Characteristic Problem. 
Integrating, 
\be
x = t + u            \m , 
\label{1-1}
\ee 
\be
y = t                \m , 
\label{1-2}
\ee 
\be
\sbiQ = \sbiQ(u)     \m .   
\label{1-3}
\ee 
Next, eliminating $t$ between (\ref{1-1}-\ref{1-2})
\be 
u = x - y            \m .  
\label{Char-1}
\ee 
Finally, substituting (\ref{Char-1}) in (\ref{1-3}), (\ref{P(x-y)}) is recovered. 

\m 

\n For $N = 3$, 
\be 
( \pa_x + \pa_y + \pa_z ) \sbiQ = 0                        \m .  
\label{Tr-PE-2a}
\ee 
This is a single PDE, so its characteristics can be found from 
\be 
\frac{\d x}{1}  \es  \frac{\d y}{1}  \es  \frac{\d z}{1}  \m .
\ee 
While direct integration is now not possible all in one go, by the flow method (\ref{Tr-PE-2a}) is equivalent to the ODE system
\be 
\dot{x} = 1      \m , 
\ee 
\be 
\dot{y} = 1      \m , 
\ee
\be 
\dot{z} = 1      \m , 
\ee
\be 
\dot{\sbiQ} = 0   \m , 
\ee
to be solved as a Free Characteristic Problem. 
Integrating, 
\be
x = t + u            \m , 
\label{2-1}
\ee 
\be
y = t + v            \m , 
\label{2-2}
\ee 
\be
z = t                \m , 
\label{2-3}
\ee 
\be
\sbiQ = \sbiQ(u, \, v)    \m .   
\label{2-4}
\ee 
Next, eliminating $t$ from (\ref{2-3}) in (\ref{2-1}-\ref{2-2}) gives the form taken by the characteristic coordinates, 
\be 
u = x - z            \m , 
\ee 
\be 
v = y - z            \m .  
\ee 
Finally, substituting these in (\ref{2-4}) yields the preserved quantities, 
\be
\sbiQ = \sbiQ(x - z,  \, y - z)  \m .
\ee 
{\bf Remark 1} $x - y$ could as well feature among these; what we have is a {\sl basis} of relative separations; 
\be 
x - y = (x - z) - (y - z)    \m , 
\ee 
so this functional dependence is already included by linear dependence.  

\m 

\n The arbitrary-$N$ case is no harder to solve. 
Its preserved equation PDE (\ref{Tr-PE}) is equivalent by the flow method to the ODE system 
\be 
\dot{q}_I = 1      \m , 
\ee 
\be 
\dot{\sbiQ} = 0     \m , 
\ee
to be solved as a Free Characteristic Problem. 
Integrating, 
\be
q^i = t \, 1^i  + u^i    \m , 
\label{qit}
\ee  
\be 
q_N = t                  \m , 
\label{qNt}
\ee 
\be
\sbiQ = \sbiQ(u^i)         \m .   
\label{Pui}
\ee 
The $i$-index here runs from $1$ to $n := N - 1$, and $1^i$ is the $n$-vector of 1's. 
So we now have an $n$-vector of characteristics $u^i$ and we are treating $q_N$ differently from the other $q^i$ (again an arbitrary choice corresponding to a picking a basis).  
Next, eliminating $t$ using (\ref{qNt}) in (\ref{qit}) gives the characteristic coordinates to be  
\be 
u^i = q^i - q^N =: r^{Ni}                 \m , 
\label{ui}
\ee 
the notation $r^{Ni}$ referring to the relative point separation between $q^N$ and $q^i$.   
Finally substituting (\ref{ui}) in (\ref{Pui}), the preserved quantities are 
\be
\sbiQ  \es  \sbiQ \left( q^i - q^N \right)  
      \es  \sbiQ \left( r^{iN} \right)                   \m .
\label{PqiqN}
\ee 
For $x_i$ a basis, so is the arbitrary linear combination $y^i = {M^i}_jx^j$ for $\u{\u{M}}$ invertible. 
In particular, subsequent sections benefit from rephrasing (\ref{PqiqN}) by use of the particular linear combination 
\be 
\sbiQ  \es  \sbiQ(\rho^i)                   \m , 
\ee 
These $\rho^i$ are some $N$-point choice of (if relevant, mass-weighted) {\it relative Jacobi coordinates} \cite{Marchal, I, Minimal-N}.
Such bases are not uniquely determined, but any choice thereof will do for the purposes of the current Series.  

\m 

\n We use ${\bm{-}}$ as shorthand for these relative separation {\sl differences}, by which we summarize our answer as the preserved quantities being of the form 
\be 
\sbiQ  \es  \sbiQ(\, {\bm{-}} \,)                     \m : 
\ee
suitably-smooth functions of differences.

\section{1-$d$ dilations}

On $\mathbb{R}^d$, the dilations form the automorphism group 
\be 
Dil = \mathbb{R}_+ \m,  
\ee
independently of dimension $d$. 

\m 

\n This case has a single preserved equation, now  

\n\be 
\biq \circ \bnabla \sbiQ  \:=  \sum_{I = 1}^N q^I \pa_I \sbiQ  \:= \sum_{I = 1}^N q^I \frac{\pa \sbiQ}{\pa q^I}  \es  0  \m , 
\label{Dil-PE}
\ee
where bold font and $\circ$ denote configuration space vectors and inner products respectively, and the second equality is specific to 1-$d$.  

\m 

\n For $N = 1$, this simplifies to the ODE 
\be 
x \, \frac{\d \sbiQ}{\d x} \es 0                       \m . 
\ee 
For $x \neq 0$, this reduces to (\ref{Tr-PE(1,1)}), and so is solved by the trivial $x = const$. 

\m 

\n While $x = 0$ permits $\sbiQ$ arbitrary, $x = 0$ is too rigid a restriction for this to entail further functional dependence for preserved quantities.  

\m 

\n $N = 2$ is minimal for (\ref{Tr-PE}) to be a PDE,  
\be 
(x \, \pa_x + y \, \pa_y)\sbiQ = 0                         \m ,  
\label{Dil-PE-2}
\ee 
and is also minimal for there to be nontrivial preserved quantities  
(One can `count out' to establish the minimality for a given $\bFrM$ and $Aut(\bFrM, \bsigma)$ \cite{Minimal-N, Minimal-N-2}.)    

\m 

\n Being a single PDE, (\ref{Dil-PE-2}) is equivalent by the flow method to the ODE system
\be 
\dot{x} = x      \m , 
\ee 
\be 
\dot{y} = y      \m , 
\ee
\be 
\dot{\sbiQ} = 0  \m , 
\ee
to be solved as a Free Characteristic Problem. 
Integrating, 
\be
x = u \, \mbox{exp}(t)  \m , 
\label{W-1}
\ee 
\be
y = \mbox{exp}(t)       \m , 
\label{W-2}
\ee 
\be
\sbiQ = \sbiQ(u)          \m .   
\label{W-3}
\ee 
Next, eliminating $t$ between (\ref{W-1}) and (\ref{W-2})
\be 
u \es \frac{x}{y}       \m , 
\ee 
so the characteristic variable is a {\it ratio} in the original coordinates. 
Finally, substituting this in (\ref{W-3}) gives the preserved quantities 
\be 
\sbiQ \es \sbiQ \left(  \frac{x}{y}  \right) \m . 
\ee 
We moreover recognize (\ref{Dil-PE}) as the {\sl Euler homogeneity equation of degree zero}, in $N$ independent variables.
It is well-known that this is solved by the functions of the $n$ independent ratios supported by the $N$ original independent variables.  
Without loss of generality,  we can pick 
\be 
{\cal Q}^i \:= \frac{q^i}{q^N}
\ee 
to be a basis for these $n$ ratios, and so 
\be 
\sbiQ = \sbiQ({\cal Q}^i)                   \m .
\label{P-R}
\ee 
We use ${\bm{/}}$ as shorthand for these ratios, by which we summarize our answer as the preserved quantities being of the form 
\be 
\sbiQ = \sbiQ(\, {\bm{/}} \,)  \m :  
\ee
suitably-smooth functions of ratios. 

\m 
 
\n Let us finally show how to recover this general-$N$ case's preserved quantities solution by recasting (\ref{Dil-PE}) by the flow method as the ODE system 
\be 
\dot{q}^I = q^I  \m , 
\ee 
\be 
\dot{\sbiQ} = 0   \m , 
\ee
to be solved as a Free Characteristic Problem. 
Integrating,  
\be
q^i = u^i  \, \mbox{exp}(t)               \m , 
\label{X-1}
\ee  
\be 
q_N = t                                   \m ,
\label{X-2}
\ee 
\be
\sbiQ = \sbiQ(u^i)                          \m .   
\label{X-3}
\ee 
Next eliminating $t$ from (\ref{X-2}) in (\ref{X-1}),   
\be 
u^i  \es  \frac{q^i}{q_N} = {\cal Q}^{i}  \m .  
\ee 
Thus, finally substituting this in (\ref{X-3}), (\ref{P-R}) is derived.\footnote{In the context of a general Euler homogeneity equation, 
this flow approach features in e.g.\ \cite{John}.}

\section{1-$d$ dilatations}

In $\mathbb{R}$, the dilatations form the automorphism group 
\be 
Dilatat(1)  \es  Tr(1) \rtimes Dil  
            \es  \mathbb{R} \rtimes \mathbb{R}_+  \m ,   
\ee 
where $\rtimes$ denotes semidirect product \cite{Cohn}. 

\m 

\n This case gives our first example of a preserved equation {\sl system}: the two PDEs  

\n\be 
\sum_{I = 1}^N \pa_I \sbiQ      \es  0     \m , 
\label{Dilatat-PE-1}
\ee 

\n\be 
\sum_{I = 1}^N q^I \pa_I \sbiQ  \es  0    \m .
\label{Dilatat-PE-2}
\ee
Counting out, the minimal case to support a nontrivial solution is $N = 3$. 
Our first approach is to solve each of these equations piecemeal, as per the previous two sections. 
We then require the {\sl compatibility equation} 
\be 
\sbiQ(x - z, \, y - z)  \es  \sbiQ\left( \frac{x}{z} \, , \, \, \frac{y}{z} \right)  \m .
\ee 
This is a functional dependence equation.
This particular functional dependence equation is solved by 
\be 
\sbiQ   \es  \sbiQ\left(  \frac{x - z}{y - z}  \right)                                                      \m , 
\ee 
since aside from being manifestly a function of $x - z$ and $y - z$, 
\be 
f(x - z, \, y - z \m \mbox{alone})  \es  \frac{x - z}{y - z}  
                                 \es  \frac{\frac{\mbox{\normalsize $x$}}{\mbox{\normalsize $z$}} - 1}{\frac{\mbox{\normalsize $y$}}{\mbox{\normalsize $z$}} - 1}  
                                 \es  f\left( \frac{x}{z} \, , \, \, \frac{y}{z} \m \mbox{alone} \right)       \m . 
\ee 
\n This analysis moreover generalizes to the arbitrary-$N$ case. 
Our compatibility equation is now the {\sl dilatational functional dependence equation}, 
\be 
\sbiQ(\rho^i)  \es  \sbiQ({\cal Q}^i)                     \m .
\ee 
This is solved by 
\be 
\sbiQ  \es  \sbiQ(\Frr^{\barr}) 
\ee 
for the basis of ratios of relative differences 
\be
\Frr^{\barr} \:= \frac{r^{\barr N}}{r^{n N}}   \m . 
\ee 
$\bar{r}$ here takes the values 1 to $\bar{n} := n - 1 = N - 2$. 
This illustrates a {\it composition principle}: imposing just translations preserves difference $-$, imposing just dilations gives ratios ${\bm{/}}$, 
but imposing both yields {\sl ratios of differences}, ${\bm{-/-}}$.

\m 

\n We may thus summarize our answer as dilatational preserved quantities being of the form 
\be 
\sbiQ  \es  \sbiQ(\, {\bm{- / -}} \,)  \m :
\ee
suitably-smooth functions of ratios of differences. 

\m 

\n{\bf Remark 1} There is moreover a way of avoiding the compatibility equation in this case. 
Solve (\ref{Dilatat-PE-1}) first.
Then, working with the $\rho^i$, it so happens that (\ref{Dilatat-PE-2}) takes the form 

\n\be 
\sum_{i = 1}^n \rho^i\pa_{\rho^i}  \sbiQ   \es  0  \m . 
\label{Dilatat-PE-3}
\ee
This (\ref{Dilatat-PE-1}) is just like (\ref{Dilatat-PE-2}) except that it is summed over one object less.  
This coincidence is in fact constructed -- and the first of many -- by taking Jacobi's special linear combination. 
This works because Jacobi coordinates keep the form $M_{IJ}$ -- as features in $q^I \pa_I = q^I {M_I}^J \pa_J$ -- diagonal upon passing to relative separations. 
This does not automatically happen because taking a basis of separations -- such as $x - z$ and $y - z$ for $N = 3$ -- 
then features $(x - y)^2$ on an even footing in the transformed form. 
So upon expressing this term in our basis, 
\be 
(x - y)^2 = (x - z - (y - z))^2 \m \m \mbox{ containing } \m  \m  (x - z)(y - z) \m \mbox{ cross terms }
\ee 
rendering it nondiagonal. 
The problem of diagonalizing this form yields Jacobi coordinates. 
This is at the slight cost of expanding ones conceptualization from relative point separations 
to relative {\sl point-cluster} separations between point subsystem centres of mass rather than necessarily just points. 
See \cite{I, Minimal-N} for recent reviews with specific examples and discussion.

\m 

\n Now the transformation 

\n\be 
q_I \m \longrightarrow \m (\rho^i, R)   \m ,
\ee 
for centre-of-mass position 

\n\be 
R \:= \frac{1}{N}\sum_{I = 1}^N q^I  
\ee 
places (\ref{Dilatat-PE-1}) in the form 

\n\be 
\left(  \sum_{i = 1}^n \rho^i\pa_{\rho^i} + R \, \pa_R  \right) \sbiQ  \es  0  \m . 
\ee    
But, having solved (\ref{Dilatat-PE-1}) $\sbiQ$ is a function of relative separations alone, so the last term returns zero, and can thus be dropped.  

\m 

\n Involving one object less moreover makes no difference to how the general-$N$ PDE is solved, the previous section's working returning immediately 
\be 
\sbiQ  \es  \sbiQ \left(  \frac{\rho^i}{\rho^j}  \right) 
      \es  \sbiQ({\cal R}^{\barr})  
	  \es  \sbiQ(\, {\bm{-/-}} \,)                               \m .  
\ee 
We term this working the {\it sequential method based on passing to centre of mass coordinates}; 
it is possible because of the simple semidirect product structure of the underlying automorphism group.

\section{Arbitrary-$d$ translations}

On $\mathbb{R}^d$, the translations form the automorphism group 
\be 
Tr(d) = \mathbb{R}^d \m . 
\ee 
The preserved equations are now the system of $d$ equations   

\n\be 
\sum_{I = 1}^N \, \u{\nabla}_I \sbiQ  \es  0  \m , 
\label{Tr-PE-N}
\ee 
the underlines denoting spatial vectors, so $\u{\nabla}_I$ is the gradient operator with respect to $q^I$, for each value of $I$.    

\m 

\n Counting out, $N = 2$ is minimal for (\ref{Tr-PE-N}) to support nontrivial solutions.  
(\ref{Tr-PE-N}) moreover admits blockwise decomposition as follows.  

\m 

\n Consider first the 2-$d$ case. 
Then two separate uses of Sec 3's working, hanging 1 and 2 indices upon them and allowing for 1's solution to depend unrestrictedly on 2's variables, 
gives the compatibility equation

\be 
\sbiQ(x_1 - y_1, \, x_2, \, y_2)  \es  \sbiQ(x_1, \, y_1, \, x_2 - y_2) \m .
\ee
This is moreover immediately solvable by superposition, 
\be
\sbiQ  \es  \sbiQ(x_1 - y_1, \, x_2 - y_2)                     \m , 
\ee   
furthermore admitting the vectorial repackaging 
\be 
\sbiQ  \es  \sbiQ(\u{x} - \u{y})                                \m .
\ee 
The availability of this method is underlied by the direct-product decomposition 
\be 
Tr(2) \es  \mathbb{R}^2  \es  \mathbb{R} \times \mathbb{R}  \es  Tr(1) \times Tr(1) 
\ee 
and the corresponding component-by-component equations being totally decoupled from each other, 
so each can be solved in terms of its own variable while leaving all the other equations' variables as free functions.  

\m 

\n This description moreover extends by a basic induction to solving the $Tr(d)$ case as a direct product with $d$ $Tr(1)$ factors, and with arbitrary $N$ as well, yielding   
\be 
\sbiQ  \es  \sbiQ(\u{r}^i)  
      \es  \sbiQ(\u{\rho}^i)  
	  \es  \sbiQ(\, {\bm{-}} \,)                             \m .
\ee 
{\bf Remark 1} This blockwise-and-superposition method is furtherly useful for direct-product automorphism groups 
\be 
Aut = \bigtimes_{\alpha = 1}^k A_{\alpha} \m , 
\ee
in which the preservation equations of each block $A_{\alpha}$ involve just on the same block's variables.

\section{Arbitrary-$d$ dilations and dilatations}

For arbitrary-$d$ dilations, the single preserved equation is still of the previously given form 
\be 
\biq \circ \bnabla \sbiQ = 0                              \m ,
\label{Dil-PE-3}
\ee
but now signifying 

\n\be 
\sum_{I = 1}^N q^{Ia} \frac{\pa \sbiQ}{\pa q^{Ia}} \es 0  \m , 
\label{Dil-PE-4}  
\ee 
since configuration vectors are now $\mathbb{R}^{Nd}$ vectors rather than just $\mathbb{R}^N$ ones.  
(\ref{Dil-PE-4}) is clearly still a single equation, and of Euler homogeneity of degree zero type.  
So we know our system's $N \, d$ independent variables supports $N \, d - 1$ independent ratios, without loss of generality 
\be 
{\cal Q}^{\Delta} \:= \frac{q^{Ia}}{q^{Nd}}              \m . 
\ee 
and that 
\be 
\sbiQ  \es  \sbiQ({\cal Q}^{\Delta})  
      \es  \sbiQ(\, {\bm{-}} \,)
\ee 
is the solution for the dilational preserved quantities. 

\m 

\n{\bf Remark 1} There is moreover an insight to be gleaned from the $> 1$-$d$ version of this example. 
Namely that passing to multi-spherical-polar coordinates -- $N$ copies of $d$-dimensional spherical polar coordinates -- $(r^I, \Omega^{I\barr})$ sends (\ref{Dil-PE-4}) to just 

\n\be 
\sum_{I = 1}^N r^{I} \frac{\pa \sbiQ}{\pa r^{I}}  \es  0  \m . 
\label{Dil-PE-5}  
\ee
But this is just the 1-$d$ dilational equation again, solved by $n$ ratios
\be 
{\cal S}^i \es \frac{r^i}{r^N} \m :
\ee
\be 
\sbiQ  \es  \sbiQ({\cal S}^i, \Omega^{I \barr}) 
      \es  \sbiQ \left (\frac{r^i}{r^N}, \Omega^{I\barr} \right)  
	  \es  \sbiQ(\, {\bm{/}} \,)                                     \m .
\ee
\n{\bf Remark 2} The spherical angles $\Omega$ enter this working as unrestricted functions rather than characteristics. 
This amounts to splitting the ratios into pure-length-ratios and angles, as well as explaining that the dilational operators act on only the former.   

\m 

\n For $d$-dimensional dilatations
\be 
Dilatat(d)  \es  \mathbb{R}^d \rtimes \mathbb{R}_+  \m , 
\ee 
the preserved equations system consists of the $d + 1$ PDEs  

\n\be 
\sum_{I = 1}^N \u{\nabla}_I \sbiQ \es 0                        \m ,   
\label{Tr-PE-N2}
\ee 
\be 
\biq \circ \bnabla \sbiQ \es 0                                 \m . 
\label{Dil-PE6}  
\ee
The `centre of mass' sequential method applies, yielding 

\n\be 
\brho \circ \bnabla_{\rho} \, \sbiQ  \es \sum_{i = 1}^n \rho^{ia} \pa_{\rho^{ia}} \sbiQ  \es 0  \m  
\label{Dil-PE7}  
\ee
(with $\circ$ now being the $\mathbb{R}^{dn}$ configuration space's scalar product).
This is solved by
\be 
\sbiQ = \sbiQ(\Frr^{\Delta})                                     \m ,
\ee 
or, using relative Jacobi multipolar coordinates, 
\be 
\sbiQ = \sbiQ({\cal R}^{\Delta})                                 \m .
\ee 
In each case, $\Delta$ form a basis of $n \, d - 1$ ratios.  
Both forms can be summarized by 
\be 
\sbiQ  \es  \sbiQ( \, {\bm{-/-}} \, )                              \m .
\ee
\n{\bf Remark 3} The dilationally-invariant configuration space is 
\be 
\FrD(d, N)  \:=  \frac{\FrQ(d, N)}{Dil}  
            \es  \frac{\mathbb{R}^{Nd}}{\mathbb{R}_+}  
			\es  \mathbb{S}^{Nd - 1}                     \m ,  
\ee
both metrically and topologically \cite{AMech, I}.

\m 

\n{\bf Remark 4} The dilatationally-invariant configuration space is 
\be 
\FrP(d, N)  \:=  \frac{\FrQ(d, N)}{Dilatat(d)}  
            \es  \frac{\mathbb{R}^{Nd}}{Tr(d) \rtimes \mathbb{R}_+} 
			\es  \frac{\mathbb{R}^{nd}}{\mathbb{R}_+} 
			\es  \mathbb{S}^{nd - 1}                     \m ,  
\ee
both metrically and topologically. 
This is known as Kendall's {\it preshape space} \cite{Kendall84, Kendall}.  

\m 

\n{\bf Remark 5} Functions thereover can thus be interpreted as functions of coordinates on spheres. 
In this context, the ratios used above receive the interpretation of {\it Beltrami coordinates}. 
One can moreover pass from this `more projective' formulation to the `more spherical' formulation in terms of hyperspherical angles.
So, firstly, for dilational preserved quantities, 
\be 
\sbiQ = \sbiQ(\btheta) 
\ee 
where $\btheta$ are $\mathbb{S}^{Nd - 1}$-hyperspherical coordinates. 
Secondly, for dilatational preserved quantities, 
\be 
\sbiQ = \sbiQ(\bTheta) 
\ee 
where $\bTheta$ are $\mathbb{S}^{nd - 1}$-hyperspherical coordinates,
This is the first of a number of illustrations in the current paper that 
{\sl mastery of the topology and geometry of the configuration space enhances understanding of the function spaces thereover of preserved quantities}.

\section{2-$d$ rotations and Euclidean transformations}

2-$d$ rotations form the automorphism group 
\be  
Rot(2) = SO(2)  \m .
\ee 
The corresponding preserved equation is   
\be 
(\biq \cr \bnabla)_{_{\mbox{\scriptsize $\perp$}}} \sbiQ  = 0  \m .  
\label{Rot-PE-1}
\ee
For $N = 1$, this becomes  
\be 
(x \, \pa_y - y \, \pa_x)\sbiQ = 0  \m .
\ee 
Under passage to plane polar coordinates $(r, \phi)$, the reformulation  
\be 
\pa_{\phi}\sbiQ = 0    
\ee
is well-known.
Integrating, 
\be 
\sbiQ = \sbiQ(r)  \m : 
\ee 
functions of radius alone. 

\m 

\n The arbitrary-$N$ case is a straightforward extension to set up, now using multipolar coordinates, 

\n\be 
\sum_{I = 1}^N\pa_{\phi^I} \sbiQ  \es  0  \m .   
\ee
By the flow method, this is equivalent to the ODE system 
\be 
\dot{\phi}^I = 1  \m ,
\ee 
\be 
\sbiQ = 0          \m ,
\ee 
to be solved as a Free Characteristic Problem.    
Integrating, 
\be 
\dot{\phi}^i = t \, 1^i + u^i  \m , 
\label{Rot-1}
\ee
\be 
\dot{\phi}_N = t               \m , 
\label{Rot-2}
\ee
\be 
\sbiQ = \sbiQ(u^i)               \m .
\label{P=P(u)}
\ee
Thus, eliminating $t$ using (\ref{Rot-2}) in (\ref{Rot-1}),
\be 
u^i = \phi^i - \phi^N =: \varphi^i \m 
\ee
so $N \geq 2$'s models' new feature is acquisition of angular dependence.
Finally substituting into (\ref{P=P(u)}),   
\be 
\sbiQ  \es  \sbiQ(\varphi^i, r^I)  \es  \sbiQ( \, {\bm{ \cdot }} \, ) 
\label{P=dot}
\ee
$\mb{\cdot}$ here denotes {\it dot} alias {\it scalar product} of vectors. 
This last equality follows from 
\be  
r^I = \sqrt{\u{q}^I \cdot \u{q}^I} \m \mbox{ (no sum) }  
\ee 
-- reinterpreting radii as {\it vector magnitudes}, more precisely radii from the model's absolute origin as position vector magnitudes -- and 
\be  
\varphi^{i}  \es  \mbox{arcos} \left( \frac{\u{q}^i \cdot \u{q}^N}{||\u{q}^i|| \, ||\u{q}^N||} \right)  \m \mbox{ (no sum) } \m : 
\ee
the definition of angle applied to the absolute case between $\u{q}^i$ and $\u{q}^N$.  
Thus (\ref{P=dot}) reads that the rotational preserved quantities are suitably smooth functions of dots.  

\m 

\n Next, on $\mathbb{R}^2$, the automorphism group of isomorphisms is 
\be 
Isom(\mathbb{R}^2)  =  Eucl(2)  
                   \es  Tr(2) \rtimes Rot(2)   
				   \es  \mathbb{R}^2 \rtimes SO(2)  \m , 
\ee 
where `Eucl' stands for Euclidean. 
In this case, the preserved equations form the system of 3 PDEs 

\n\be 
\sum_{I = 1}^N \u{\nabla}_I \sbiQ = 0  \m ,
\label{solve-first}
\ee 
\be 
(\biq \cr \bnabla)_{_{\mbox{\scriptsize $\perp$}}} \sbiQ  \es 0    \m .   
\label{solve-second}
\ee 
The  `centre of mass' sequential method applies again: solve (\ref{solve-first}) to obtain 
\be 
\sbiQ  \es  \sbiQ\left(\rho^i\right) \m , 
\ee
and substitute in (\ref{solve-second}) to get the PDE
\be 
(\brho \cr \bnabla_{\rho})_{_{\mbox{\scriptsize $\perp$}}} \sbiQ \es 0    \m .   
\ee 
Recasting this in multipolar coordinates, 

\n\be 
\sum_{i = 1}^n \pa_{\theta^i} \sbiQ = 0 \m , 
\ee 
which is solved parallelling the first half of this section by 
\be 
\sbiQ \es \sbiQ(\Theta^{\bar{r}}, \rho^i) \es \sbiQ({\bm{- \cdot -}}) \m .
\ee 
This last equality follows from applying the definition of magnitude to mass-weighted relative Jacobi vectors, 
\be  
\rho^i = ||\rho^i|| = \sqrt{\u{\rho}^i \cdot \u{\rho}^i} \m , 
\ee 
and the definition of angle in the relative case between $\u{\rho}^{\barr}$ and $\u{\rho}^n$: 
\be  
\Theta^{\bar{r}}  \es  \mbox{arcos} \left( \frac{\u{\rho^{\bar{r}}} \cdot \u{\rho}^n}{||\u{\rho}^{\bar{r}}|| \, ||\u{\rho}^n||} \right)  \m . 
\ee
Thus Euclidean preserved quantites are suitably smooth functions of dots of differences.

\section{Rotation--dilation independence in 2-$d$}

Rotation and dilation are independent on $\mathbb{R}^2$, forming the combined geometrical automorphism group 
\be 
Rot(2) \times Dil  \es  SO(2) \times \mathbb{R}_+
\label{RotXDil}
\ee
The corresponding preserved equation system is 
\be 
(\biq \cr \bnabla )_{_{\mbox{\scriptsize $\perp$}}} \sbiQ \es 0    \m , 
\ee 
\be 
\biq \cdot \bnabla \sbiQ = 0  \m , 
\ee  
Counting out, $N = 2$ is minimal. 
Employing bipolar coordinates in this case, our equations reduce to 
\be 
( \pa_{\phi_1} + \pa_{\phi_2} ) \sbiQ  = 0  \m , 
\ee 
\be 
( r_1 \pa_{r_1} + r_2 \pa_{r_2} )\sbiQ  = 0  \m . 
\ee 
This choice of coordinates has succeeded in fully decoupling our system into blocks. 
The possibility of this -- {\it factorizability} -- is underlied by the direct product structure of (\ref{RotXDil}).  
Solving piecemeal as per Secs 11 and 12 gives the compatibility equation  
\be 
\sbiQ(  \varphi, \, r_1, \, r_2)  \es  \sbiQ\left(\phi_1, \, \phi_2, \, \frac{r_1}{r_2} \right)         \m , 
\ee 
which moreover superposes to give preserved quantities 
\be 
\sbiQ \es \sbiQ \left( \varphi, \, \frac{r_1}{r_2} \right) \es \sbiQ( \, {\bm{ \cdot / \cdot }} \, )  \m .  
\ee 
Using multipolar coordinates, this method extends to the $N$-point case, giving the preserved-equation system 

\n\be 
\sum_{I = 1}^N \pa_{\phi^I} \sbiQ = 0  \m , 
\ee 

\n\be 
\sum_{I = 1}^N r^I\pa_{r^I} \sbiQ = 0  \m , 
\ee 
which is solved by 
\be 
\sbiQ \es \sbiQ \left( \varphi^{i}, \, \frac{r^{i}}{r^N} \right) \es \sbiQ( \, {\bm{\cdot / \cdot}} \, )  \m .  
\ee 
Thus rotational-and-dilational preserved quantities are suitably smooth functions of ratios of dots.

\section{2-$d$ similarity group}

\n On $\mathbb{R}^2$, the similarity group of automorphisms is 
\be 
Sim(2)             \es  Tr(2) \rtimes ( Rot(2) \times Dil )  
				   \es  \mathbb{R}^2 \rtimes ( SO(2) \times \mathbb{R}_+)
\ee
The corresponding preserved equation system consists of the 4 PDEs 

\n\be 
\sum_{I = 1}^N \u{\nabla}_I \sbiQ  \es 0  \m ,
\label{solve-first-2}
\ee 

\n\be 
\biq \cdot \bnabla \sbiQ = 0  \m , 
\label{solve-second-a}
\ee 

\n\be 
(\biq \cr \bnabla )_{_{\mbox{\scriptsize $\perp$}}} \sbiQ  \es  0    \m .   
\label{solve-second-b}
\ee
This is also amenable to the `centre of mass' sequential method: solve (\ref{solve-first-2}) to obtain 
\be 
\sbiQ  \es  \sbiQ(\rho^i) \m , 
\ee
(for {\sl magnitudes} $\rho^i$) and substitute in (\ref{solve-second-a}, \ref{solve-second-b}) to arrive at the PDE system 

\n\be 
(\brho \cr \bnabla_{\rho})_{_{\mbox{\scriptsize $\perp$}}} \sbiQ  \es 0    \m .   
\ee
 
\n\be 
\brho \cdot \bnabla_{\rho} \sbiQ = 0          \m .   
\ee
Recasting this in Jacobi multipolar coordinates decouples the two equations (factorizability):  

\n\be 
\sum_{i = 1}^n \pa_{\theta^i} \sbiQ \es 0       \m , 
\ee 

\n\be 
\sum_{i = 1}^n \rho^i \pa_{\rho^i} \sbiQ \es 0  \m , 
\ee 
which system is solved parallelling the previous section to give the preserved quantities 
\be 
\sbiQ \es \sbiQ(\Theta_{\bar{r}}, \, {\cal R}^{\bar{r}}) \es \sbiQ(z^{\bar{r}})  \es \sbiQ({\bm{-\cdot-/-\cdot-}})  \m .
\label{Sim-2-PQ}
\ee
Thus similarity preserved quantities are suitably-smooth functions of ratios of dots of differences. 

\m 

\n{\bf Remark 1} The penultimate form here is special to 2-$d$, wherein the ${\cal R}^{\bar{r}}$ and $\Theta^{\bar{r}}$ form complex pairs \cite{Kendall}

\n\be 
z^{\bar{r}} = {\cal R}^{\bar{r}} \, \mbox{exp}(i \, \Theta^{\bar{r}})  \m , 
\ee
which play the role of inhomogeneous coordinates on these models' shape spaces \cite{Kendall84, Kendall}, 

\n\be 
\FrS(2, N) = \mathbb{CP}^{N - 2} \m , 
\ee 
both topologically and equipped with the standard Fubini--Study metric.

\section{Arbitrary-$d$ rotations and compositions}

On $\mathbb{R}^d$, the continuous subgroups of the similarity group that involve rotations are as follows. 

\n\be 
Rot(d) \es SO(d) \m , 
\ee

\n\be
Isom(\mathbb{R}^d) \es   Eucl(d)             
                   \es  Tr(d) \rtimes Rot(d)  
				   \es  \mathbb{R}^d \rtimes  SO(d) 
\ee
rotations and dilations remain independent,  

\n\be 
Rot(d) \times Dil \m , 
\ee
and 

\n\be 
Sim(d)             \es  Tr(d) \rtimes ( Rot(d) \times Dil )  
				   \es  \mathbb{R}^d \rtimes ( SO(d) \times \mathbb{R}_+)
\ee
The rotational preserved quantities solve a $d(d - 1)/2$-dimensional system, but continue to be of the form 
\be 
\sbiQ  \es  \sbiQ(\, {\bm{\cdot}})  \m . 
\ee 
The rotational-and-dilational preserved quantities solve a $d(d - 1)/2 + 1$ dimensional system, but continue to be of the form 
\be 
\sbiQ  \es  \sbiQ(\, {\bm{\cdot / \cdot}})  \m .
\ee
The Euclidean preserved quantities solve a $d(d + 1)/2$-dimensional system, but continue to be of the form  
\be  
\sbiQ  \es  \sbiQ(\, {\bm{-\cdot-}})  \m . 
\ee
Finally, the similarity preserved quantities solve a $d(d + 1)/2 + 1$-dimensional system, but continue to be of the form  
\be 
\sbiQ  \es  \sbiQ(\, {\bm{-\cdot-/-\cdot-}})  \m .
\ee 
These results arise by extensions of the `centre of mass' sequential and factorizability methods, the latter proceeding via multi-$d$-dimensional-hyperspherical-polar coordinates.  

\m 

\n{\bf Remark 1} For $d \geq 3$, there is no longer a complex pairing of ratio and angular coordinates, 
the number of independent angular coordinates for $Sim(d)$ and $Rot(d) \times Dil$ now exceeding the number of independent ratios.
This is by $\bar{n} = N - 2$ relative ratios to 

\n\be 
N(d - 1) - \frac{d(d + 1)}{2} - 1 
\ee 
relative angles for $Sim(d)$, and $n = N - 1$ ratios to 

\n\be 
\frac{(d - 1)}{2}(2N - d)
\ee
angles for $Rot(d) \times Dil$. 

\m 

\n{\bf Remark 2} $Eucl(d)$ preserved quantities can now moreover be repackaged as $Sim(d)$'s preserved quantities, alongside 

\n\be 
\rho  :=  \sqrt{I} 
      \es \sqrt{\sum_{i = 1}^n\rho^{i \, 2}}  \m , 
\ee 
which is a {\it configuration space radius} (alias {\it hyperradius} in the Molecular Physics and Mathematical Physics \cite{MFII} literatures) 
scale variable (mass-weighted whenever appropriate), $I$ itself standing for moment of inertia about the centre of mass. 
Thus for $Eucl(d)$,
\be 
\sbiQ  \es \sbiQ({\bm{-\cdot-/-\cdot-}}, \, \rho)   \m . 
\ee 
\n{\bf Remark 3} Similarly, $Rot(d)$ preserved quantities can now be repackaged as $Rot(d) \times Dil$'s alongside 

\n\be 
\chi  :=  \sqrt{J}  
      \es \sqrt{\sum_{I = 1}^n||\u{\chi}^I||^2}   \m , 
\ee
for $\u{\chi}^I$ the if-needs-be mass-weighted counterparts of the $\u{q}^I$, and $J$ the total moment of inertia about the model's absolute origin.  

\m 

\n{\bf Remark 4} We can furthermore interpret Remark 2 in terms of the configuration space `relational space' \cite{FileR}

\n\be 
\FrR(d, N) \:= \frac{\FrQ(\mathbb{R}^d, N)}{Eucl(d)}
\ee 
being the cone over shape space: 

\n\be 
\FrR(d, N) = \mC(\FrR(d, N))  \m .
\ee 
`Cone' is here meant in both the topological and metrical senses, with $\rho$ serving as its radial coordinate. 
The functions thereover are then comprised of functions `of shape' ${\bm{-\cdot-/-\cdot-}}$ and of radius $\rho$.  

\m 

\n{\bf Remark 5} We can similarly interpret Remark 3 in terms of the configuration space `rotational space' \cite{FileR, AMech}

\n\be 
\FrR\mo\mt(d, N) \:= \frac{\FrQ(\mathbb{R}^d, N)}{Rot(d)}
\ee 
being the cone over dilational-and-rotational space \cite{AMech}: 

\n\be 
\FrD\FrR\mo\mt(d, N) = \mC(\FrR\mo\mt(d, N))  \m .
\ee
The functions thereover are then comprised of dilationally-and-rotationally invariant functions with respect to a fixed absolute origin, ${\bm{\cdot/\cdot}}$ , and of radius $\chi$.  

\m 

\n{\bf Remark 6} For $d = 1$, the $Sim(1) = Dilatat(1)$ shape spaces take the form of spheres by Remarks 1 and 2 of Sec 11, 
so the corresponding Euclidean relational spaces can furthermore be expressed as \cite{FileR} 

\n\be 
\FrR(1, N) = \mC(\FrS(1, N)) = \mC(\mathbb{S}^{\bar{n}}) = \mathbb{R}^n \m . 
\ee 
On the other hand, for $d = 2$, the corresponding Euclidean spaces can furthermore be expressed as \cite{Quad-I}

\n\be 
\FrR(2, N) = \mC(\FrS(2, N)) = \mC(\mathbb{CP}^{\bar{n}})  \m , 
\ee 
by which the functions thereover can be re-expressed as 

\n\be 
\sbiQ = \sbiQ(z^{\bar{r}}, \, \rho)                            \m .
\ee 

\vspace{10in}

\section{Conclusion}

\n We have considered preserved equations, solution of which gives systematically a given geometry's preserved quantities.
Ab initio, these equations take the form of zero Lie brackets with the sum-over-points of generators. 

\m 

\n We moreover recast these preserved equations as PDEs. 
These are first-order, linear in the $\sbiQ$ and homogeneous in the first-order derivatives -- all simplifying features -- 
but are can also be over-determined systems:                                                     complicating features.

\m 

\n These equations additionally coincide with a subset of observables equations via our Bridge Theorem.  
These start off as zero Poisson brackets with constraints, whether strongly or weakly so, in the sense of Dirac \cite{Dirac}.   
In each case, we need algebraic closure beforehand, whether of constraints or of sums-over-points of generators.
Each notion of observables or preserved quantity itself moreover then brackets-closes.
The notions of preserved quantity a given geometry supports form a bounded lattice, dual to that of subalgebraic structures of the sums-over-points of generators.
Similarly the notions of observables that a given physical model supports form a bounded lattice, dual to that of subalgebraic structures of the constraints.
We obtain the following results for preserved quantities and observables.  
 
\m 

\n 1) In the generic case, in the sense of a geometry possessing no generalized Killing vectors --  i.e.\ continuous geometrical automorphisms -- 
preserved quantities are just free functions over the geometry.  
Similarly, in the case of a physical theory with no constraints, the classical observables are just free functions over the phase space. 
(Differential Geometry and the differential equations of Physics do restrict these functions to be reasonably smooth).  

\m 

\n 2) Homogeneous PDEs admit the trivial solution; for those homogeneous-linear in the first derivatives, the trivial solution takes the form of the constant solution.  

\m 

\n 3) We argued moreover that preserved equations and observables equations are to be treated as Free alias Natural Characteristic Problems.  				

\m 

\n 4) The next most generic case is that of a single PDE, which admits a standard flow method by which one passes to an equivalent ODE system.  

\m 

\n 5) The case of a system of preservation or observables equations is not however standard.  
Overdetermination in this case does not pose a problem as regards nonexistence of nontrivial solutions, by the integrability we prove. 
This follows from Frobenius' Theorem, and is a consequence of the underlying geometrical automorphisms closing.

\m

\n For $Sim(d)$ and its geometrical subgroups, 
aside from a few cases in which there is but a single preserved equation which is amenable to a standard flow method, 
the following methods work for the ensuing preserved equation systems.

\m 

\n Method 1) Passing to the centre of mass frame induces a {\it sequential} solving step which inter-relates pairs of subgroups with and without translations, 
by recasting those with translations in the form of those without translations for one point less.

\m 

\n Method 2) The direct product rendering rotations and dilations independent supports a {\it blockwise} resolution of the corresponding preserved equations. 
By this, rotations and dilations fix disjoint sets of characteristic coordinates, by which these solutions can readily be superposed 
in case in which both rotations and dilations are among the symmetries.  

\m 

\n While Method 2) is an example of finding blockwise solvability due to {\it factorizability} entailed by a direct product in the underlying automorphism group,  
         Method 1) follows from how translations are appended by a straightforward semidirect product. 

\m 

\n We additionally provided a compact notation ${\bm{\sim}}$, ${\bm{-}}$, ${\bm{/}}$, ${\bm{\cdot}}$ for the outcome of solving the preserved equations for $Sim(d)$ and its subgroups, 
which we use in the dual bounded lattice summary of Fig \ref{Sim-Latt-2}.  
We also use the 1-$d$ case of this -- $Sim(1) = Dilatat(1)$ and its subgroups -- as an example of presheaves of automorphism groups and of preserved quantities, 
as displayed in Fig \ref{Sim-Sheaf-1-d}.
%
{            \begin{figure}[!ht]
\centering
\includegraphics[width=0.85\textwidth]{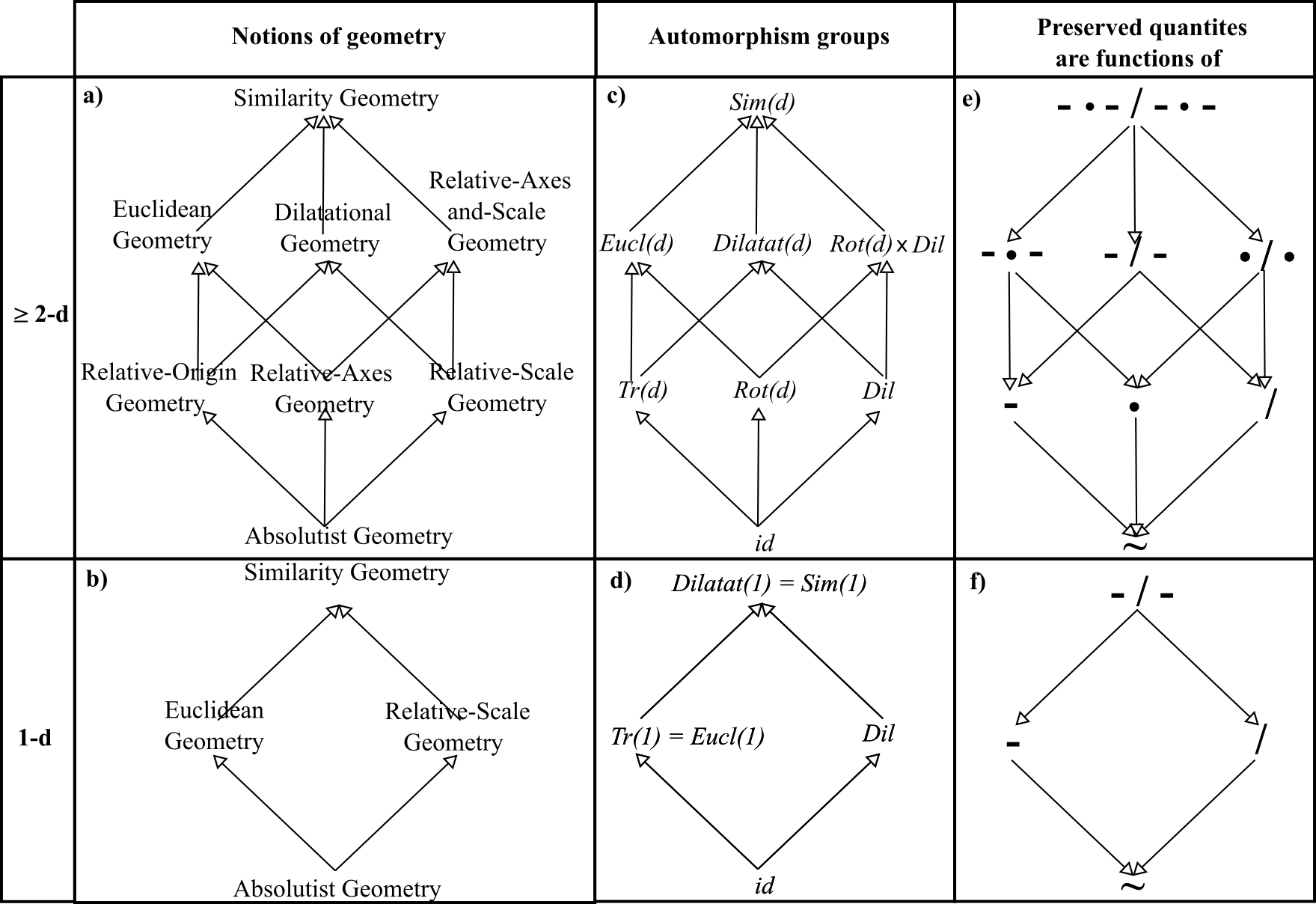}
\caption[Text der im Bilderverzeichnis auftaucht]{        \footnotesize{We here extend Fig 1 to include the dual bounded lattice of preserved quantities, 
comprising of suitably smooth functions of the indicated functional dependencies.}}
\label{Sim-Latt-2} \end{figure}          }
%
{            \begin{figure}[!ht]
\centering
\includegraphics[width=0.75\textwidth]{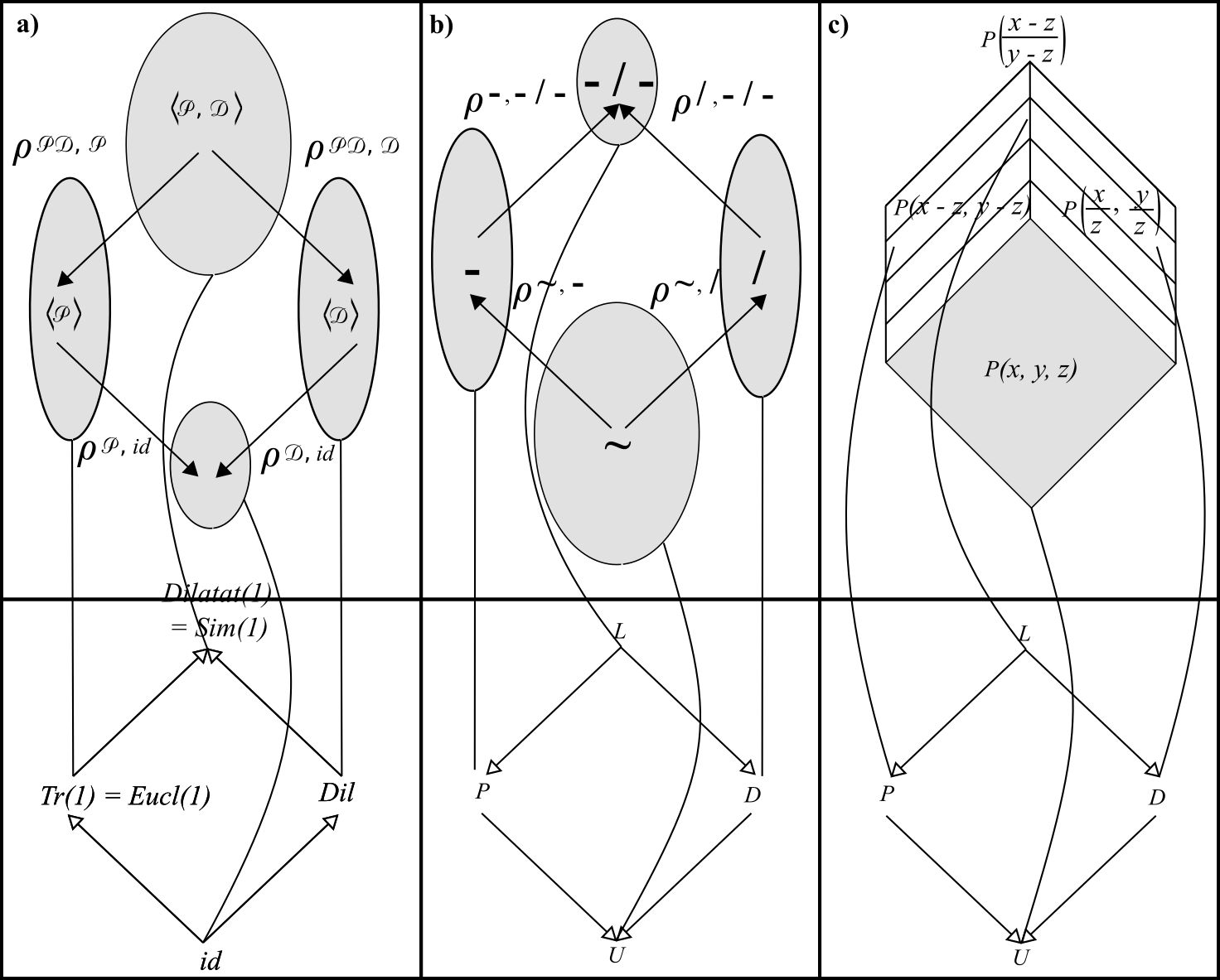}
\caption[Text der im Bilderverzeichnis auftaucht]{        \footnotesize{a) Presheaf of subalgebras over the bounded lattice of notions of subalgebra for $Sim(1)$
The solid black arrowheads denote restriction maps $\rho$, and the solid angular brackets denote span. 
b) Presheaf of preserved-quantity function-space subalgebras over the bounded dual lattice of notions of preserved quantity (or of the corresponding quotient spaces \cite{AObs4}).   
c) For the case of $N = 3$ points, this presheaf consists of the smooth functions $\tbiQ$ over the 1-$d$ line depicted vertically, where the two 2-$d$ faces, 
each spanned by the further characteristic lines running along it, and the 3-$d$ solid grey slab that these are faces of.  
Note throughout that the presheaves' restriction maps run in the opposite sense to the underlying lattice's arrows, by which these are all {\it contravariant} sheaves.}}
\label{Sim-Sheaf-1-d} \end{figure}          }

\m 

\n Some further research topics opened up by the current Article are as follows.  

\m 

\n{\bf Frontier A} Its methods of solution extends rather beyond its range of examples.
This is firstly attested by the rest of the current Series' solving the Affine, Projective and Conformal Geometry counterparts.  
The above combination of methods continues to work in the case of the affine group and its subgroups, supplemented by a compatibility equation or chain-rule sequential method.  
Method 1) however ceases to work for projective and conformal subgroups (Articles II to V), since translations are here more intricately involved than by a semidirect product.

\m 

\n A number of `non-classical' geometries follow moreover from further subgroups of Projective and Conformal Geometries' automorphisms, 
for which the current Series' direct approach to computing preserved quantities provides particular interest. 
So far, it is known that these possess standard automorphism groups but represented by non-standard combinations of generators -- 
special-conformal or special-projective transformation generators taking the place of translation generators. 
The current Series adds to this moreover that the corresponding preserved quantities are different from standard geometries' as well, 
so there are more types of geometrically-significant preserved quantities than usually enters consideration in the literature.  

\m 

\n{\bf Frontier B} \cite{AObs4} further exploits these methods to find specifically-physical Kucha\v{r} and Dirac observables for $Sim(d)$-invariant Mechanics.
This is moreover but `a tip of any Iceberg' as regards using my concrete theory of what observables are (\cite{ABeables, AObs2, ABook} and the current Article) 
and how to solve for them (as per \cite{AObs3, ABook} and especially the current Article).

\m 

\n{\bf Frontier C (Application to Foundations of Geometry)} Geometry arises from a number of different foundational standpoints 
-- e.g.\ referred to as `Pillars of Geometry' in Stillwell's account \cite{Stillwell} -- 
and the current Series is part of an expansion of this list by various further foundational schemes. 
In Stillwell's account, Klein's Erlangen Program \cite{Klein} features as Pillar 3, 
to Euclidean Construction \cite{Elements, Coxeter} being Pillar 1, 
Cartesian \cite{Descartes} through to Linear Algebra based \cite{Silvester} Geometry being Pillar 2, 
and ray diagrams leading in particular to Projective Geometry's \cite{Desargues, VY10, S04, Hartshorne} substantial axiomatic power \cite{HC32} being Pillar 4.
Klein's Erlangen Program becomes transformations, groups, and preserved quantities based Geometry \cite{Guggenheimer, Martin}.  

\m 

\n On the one hand, gives a further `Preserved Equations' `Pillar of Geometry' in entirely geomerical terms. 
This `Seventh' Pillar (by my reckoning \cite{8-Pillars}) provides zeroth principles to arrive at the preserved quantities end of Pillar 3 by means of solving specific PDEs: 
preserved equations.  
This is rather reminiscent of how automorphism equations alias generalized Killing equations provide zeroth principles to arrive at the automorphism groups end of Pillar 3 
by means of solving other specific PDEs (a much better known `Sixth Pillar' of Geometry).  

\m 

\n{\bf Frontier D (Further Interplay with Physics)} 
On the other hand -- this `Seventh Pillar' being {\sl inspired} by parallel developments in Physics concerning constraints and observables -- 
we also demonstrate equivalence between preserved quantities and a subset of observables: `strong configurational gauge observables'. 
This is the bridge over which the current Series' geometrical considerations cross into giving a full-blown theory of observables for Classical Physics. 
Background Independence \cite{A64, Dirac, A67, BB82, I89-I91, HT92, Kiefer04, RovelliBook, Giu06, ASoS, ABook} 
is moreover of even wider interest from both Foundational and Theoretical Physics, 
and Expression in terms of Observables is one of nine conceptual aspects \cite{Kuchar92, I93, APoT2, APoT3, ABook} of Background Independence. 
Difficulties with, and incompatibilities between, implementation of Background Independence aspects constitute moreover Problem of Time facets, which provides 
a further name under which Background Independence issues are referred to and covered in the literature \cite{Battelle, DeWitt67, Kuchar92, I93, APoT, APoT2, ABook, PoT-Lett}.  

\m 

\n Two of the other aspects of Background Independence -- Configurational Relationalism and Constraint Closure -- moreover bear close ties to Geometry's `Automorphism Equations' 
and `Brackets Closure' Pillars respectively, the latter being indeed the `Fifth Pillar' in my account. 
What was stated to be known four paragraphs back is via this `Fifth Pillar' \cite{Brackets-I, Brackets-II}.  

\m 

\n All in all, the interconnection between Foundations of Geometry and Background-Independent Physics is thereby large, and likely to be mutually fruitful to both subjects.  
  
\m 

\n{\bf Acknowledgments} I thank Chris Isham and Don Page for previous discussions.  
Enrique Alvarez, Jeremy Butterfield, Malcolm MacCallum and Reza Tavakol for support with my career.  

\vspace{10in}


\end{document}